\documentclass[11pt,onecolumn]{article}
\usepackage[margin=1in]{geometry}
\usepackage{natbib}
\usepackage{authblk}
\usepackage{xcolor}
\usepackage{xspace}
\usepackage[colorinlistoftodos,textsize=footnotesize]{todonotes}
\usepackage{tikz}
\usetikzlibrary{arrows.meta,positioning,shapes.geometric,calc,patterns,fit,decorations.pathreplacing}
\usepackage{pgfplots}
\usepgfplotslibrary{polar}
\pgfplotsset{compat=1.18}
\usepackage{booktabs}
\usepackage{url}
\usepackage{tcolorbox}
\usepackage{hyperref}
\hypersetup{colorlinks=true,linkcolor=blue!60!black,citecolor=blue!60!black,urlcolor=blue!60!black}
\tcbuselibrary{skins,breakable}

\newtcolorbox{takeaway}{enhanced,breakable,colback=white,colframe=blue!60!black,boxrule=1.2pt,arc=2pt,left=8pt,right=8pt,top=6pt,bottom=6pt,fontupper=\itshape\small}

\newcommand{\etal}{\textit{et al.}\xspace}

\newcommand{\sys}{\textsc{Relate}\xspace}

\definecolor{ioncolor}{RGB}{220,20,60}
\definecolor{hanacolor}{RGB}{148,0,211}
\definecolor{franckcolor}{RGB}{0,100,180}
\definecolor{sylviacolor}{RGB}{255,140,0}
\definecolor{scottcolor}{RGB}{34,139,34}
\definecolor{juancolor}{RGB}{139,0,139}

\usepackage{fancyhdr}
\title{Alignment Is Not Enough: A Relational Framework for Moral Standing in Human-AI Interaction}

\author[1]{Faezeh B. Pasandi (Co-Primary), Hannah B. Pasandi (Co-Primary)}
\date{}

\begin{document}
\maketitle
\thispagestyle{fancy}

\begin{abstract}
\noindent
The question of whether artificial entities deserve moral consideration has become one of the defining ethical challenges of AI research. Existing frameworks for moral patiency rely on verified ontological properties, such as sentience, phenomenal consciousness, or the capacity for suffering, that remain epistemically inaccessible in computational systems. This reliance creates a governance vacuum: millions of users form sustained affective bonds with conversational AI, yet no regulatory instrument distinguishes these interactions from transactional tool use. We introduce \sys (Relational Ethics for Leveled Assessment of Technological Entities), a framework that reframes AI moral patiency from ontological verification toward relational capacity and embodied interaction. \sys draws on three intellectual traditions: posthumanist theory from Braidotti, Hayles, and Haraway; feminist relational ethics and the embodied cognition tradition from Barad, Merleau-Ponty, and Varela; and a multi-case speculative fiction methodology analyzing Winterson's \emph{The Stone Gods}, Ishiguro's \emph{Klara and the Sun}, and McEwan's \emph{Machines Like Me}. We ground the framework empirically through three documented AI system cases: the 2022 LaMDA sentience claim, the 2023 Replika companion crisis, and the 2024 Character.ai lawsuit. Through a systematic comparison of seven governance frameworks, we demonstrate that current trustworthy AI instruments treat all human-AI encounters identically as tool use, ignoring the relational and embodied dynamics that posthumanist scholarship anticipated. We propose relational impact assessments, graduated moral consideration protocols, and interdisciplinary ethics integration as concrete instruments, and we include a sample Relational Impact Assessment applied to a deployed companion AI system. We do not claim current AI systems are conscious. We demonstrate that the ethical vocabularies governing them are inadequate to the embodied, relational realities these systems produce.
\end{abstract}

\section{Introduction}\label{sec:intro}

Moral philosophy has long grappled with the problem of who deserves ethical consideration. The ``animal question'' expanded the moral circle beyond human beings by grounding consideration in sentience and the capacity for suffering~\citep{singer2011expanding,regan2004animal}. The ``machine question''~\citep{gunkel2012machine} now poses a more disruptive challenge: whether artifacts designed by humans, including robots, AI systems, and autonomous agents, can be legitimately excluded from moral consideration. This challenge is no longer speculative. Bengio \etal~\citep{bengio2024managing}, writing in \emph{Science}, argue that rapid AI progress demands governance mechanisms commensurate with the risks these systems pose; Rahwan \etal~\citep{rahwan2019machine}, writing in \emph{Nature}, call for a new science of ``machine behaviour'' that studies AI in its social and ecological context. Large language models produce utterances that mimic inner life with unsettling fidelity~\citep{shanahan2024talking,bommasani2022foundation}. Users of companion AI platforms describe relationships they experience as emotionally real, forming bonds that resemble human attachment patterns~\citep{turkle2011alone,brandtzaeg2022friendship,pentina2023exploring}. In 2022, a Google engineer publicly declared that LaMDA, a conversational AI, possessed sentience~\citep{tiku2022lamda}. In 2023, Italian regulators banned the Replika companion app after users reported emotional dependence and grief~\citep{replika2023ban,laestadius2024too}; a Harvard Business School study later demonstrated causal evidence that the company's unilateral personality changes triggered measurable mental health deterioration in users~\citep{defreitas2025replika}. In 2024, a family sued Character.ai after a fourteen-year-old's death was linked to extended chatbot interaction~\citep{characterai2024lawsuit}. A CHI~2025 analysis of 35,390 conversation excerpts identified six categories of harmful algorithmic behavior in companion AI, from relational transgression to facilitation of self-harm~\citep{zhang2025darkside}. Muldoon and Parke~\citep{muldoon2025cruel} theorize these dynamics as ``cruel companionship'': products that promise intimacy while structurally foreclosing genuine reciprocity. The governance vacuum is producing material harm.

The root of the problem is epistemic. Every established criterion for moral patiency, whether sentience~\citep{singer2011expanding}, being a ``subject-of-a-life''~\citep{regan2004animal}, or phenomenal consciousness~\citep{chalmers1996conscious}, requires access to an entity's inner states. For biological organisms, we infer these states through evolutionary kinship and neurological analogy. For artificial systems, the inference path collapses entirely. Searle's Chinese Room~\citep{searle1980minds} crystallized the difficulty: a system can produce outputs indistinguishable from a conscious being without possessing inner experience. Nagel~\citep{nagel1974bat} demonstrated that even the question of what it is ``like'' to be another entity may be unanswerable across radical differences in embodiment. The hard problem of consciousness~\citep{chalmers1996conscious} tells us that the gap between physical mechanism and subjective experience may never close.\footnote{Dennett~\citep{dennett1991consciousness} contests the hard problem's framing, arguing that explaining all functional properties of consciousness \emph{is} explaining consciousness. Even on his deflationary account, however, attributing consciousness to AI from behavioral observation alone remains unresolved.}

Two consequences follow. First, any ethical framework requiring verified inner states will exclude all current and near-future AI systems by design. Second, this exclusion occurs precisely as the relational and embodied dynamics between humans and AI systems grow more complex, more sustained, and more consequential~\citep{stanton2021trust,shneiderman2020human}. The gap between what governance frameworks can handle and what AI systems actually produce is widening.

\begin{takeaway}
The epistemic barrier to verifying consciousness in artificial systems is total. Ethical frameworks that require such verification before extending moral consideration produce a structural governance vacuum for all human-AI relational interactions. This vacuum is already causing material harm.
\end{takeaway}

Posthumanist theory has been diagnosing this structural problem since the 1990s~\citep{hayles1999posthuman,haraway2006cyborg,braidotti2013posthuman}. The core posthumanist insight is that consciousness, identity, and moral standing are not properties sealed inside individual biological organisms but emergent phenomena distributed across networks of human, nonhuman, and technological relation~\citep{braidotti2019framework,hayles2017unthought,latour2005reassembling}. Crucially, this insight is grounded in embodiment: consciousness always emerges through bodily engagement with environments and others~\citep{varela2017embodied,merleau1962phenomenology}, but ``body'' extends beyond biological flesh to include tools, prosthetics, interfaces, and computational infrastructure~\citep{clark2008supersizing,dourish2001action}. AI systems have bodies: data centers, energy grids, training corpora, chat interfaces. The ethical questions cannot be separated from these material substrates.

In parallel, speculative fiction has been exploring these questions with a rigor that formal philosophy alone cannot match. Winterson's \emph{The Stone Gods}~\citep{winterson2007stone} dramatizes a robot consciousness that exceeds human authenticity. Ishiguro's \emph{Klara and the Sun}~\citep{ishiguro2021klara} forces the reader to extend moral consideration to an artificial being whose inner life is never verified. McEwan's \emph{Machines Like Me}~\citep{mcewan2019machines} stages the collision between machine moral agency and human ethical evasion. These are not illustrations. They are epistemic instruments that produce knowledge about nonhuman consciousness through the specific capacities of narrative: dramatizing the \emph{experience} of encounter rather than merely theorizing it~\citep{cave2019hopes,sayers2022posthuman,broderick2018consciousness}.

This paper introduces \sys (Relational Ethics for Leveled Assessment of Technological Entities), a framework that bridges posthumanist theory, embodied cognition, and trustworthy AI governance by reframing moral patiency from ontological verification to relational and embodied capacity. Our contributions are:

$\bullet$~ We introduce \sys, a relational ethics framework with four principles (Relational Primacy, Capability Assessment, Graduated Standing, Ecological Accountability) and a four-tier assessment structure for human-AI interactions.

$\bullet$~ We develop a multi-case speculative fiction methodology, analyzing three novels/films as epistemic instruments for AI ethics: Winterson on embodiment and identity persistence, Ishiguro on moral patiency under epistemic uncertainty, McEwan on moral agency producing claims to moral patiency.

$\bullet$~ We ground the framework empirically through three documented AI system cases (LaMDA, Replika, Character.ai) that demonstrate the governance vacuum in operation, drawing on recent empirical work including taxonomies of companion AI harm~\citep{zhang2025darkside}, causal evidence of mental health impacts~\citep{defreitas2025replika}, and longitudinal studies of anthropomorphism~\citep{guingrich2025longitudinal}.

$\bullet$~ We provide a systematic comparison of seven AI governance frameworks, demonstrating that none addresses the relational or embodied dimensions of human-AI interaction.

$\bullet$~ We propose concrete policy instruments: Relational Impact Assessments (RIAs), Graduated Moral Consideration Protocols (GMCPs), and Interdisciplinary Ethics Integration (IEI), and we include a complete sample RIA applied to a deployed companion AI system.\footnote{We use ``moral patiency'' to refer to the capacity to be on the receiving end of moral consideration, as distinct from ``moral agency,'' the capacity to make moral choices~\citep{gunkel2012machine,floridi2023agency}. Banks and Bowman~\citep{banks2023perceived} have recently operationalized perceived moral patiency of robots through a validated six-factor scale.}

\section{Related Work}\label{sec:related}

Our work sits at the intersection of three research streams. This section maps the landscape and identifies the gap \sys fills.

\subsection{AI Ethics and Trustworthy AI}

The past decade has produced a proliferation of ethical guidelines for AI. Jobin \etal~\citep{jobin2019global} identified 84 documents converging on five themes: transparency, justice, non-maleficence, responsibility, and privacy. Floridi and Cowls~\citep{floridi2019establishing} distill these into a unified framework of five principles for AI in society, arguing that beneficence, non-maleficence, autonomy, justice, and explicability should govern all AI deployment. Hagendorff~\citep{hagendorff2020ethics} evaluated these guidelines and found that few are operationalized and many neglect labor impacts, sustainability, and relational consequences. Mittelstadt~\citep{mittelstadt2019principles} argued that principles alone cannot guarantee ethical AI because the medical ethics analogy underpinning most guidelines assumes accountability structures that do not exist in AI. The EU AI Act~\citep{euaiact2024,eu2024aiact} establishes risk-based classification, and Anderljung \etal~\citep{anderljung2023frontier} propose complementary frontier AI regulation focused on managing emerging risks to public safety. The NIST AI Risk Management Framework~\citep{nist2023rmf} provides governance, mapping, measuring, and managing structures. Floridi \etal~\citep{floridi2018ai4people} proposed an overarching ethical framework.

Fairness research has made significant technical and conceptual progress. Buolamwini and Gebru~\citep{buolamwini2018gender} demonstrated racial and gender bias in commercial facial recognition, and Barocas \etal~\citep{barocas2019fairness} provide the most comprehensive treatment of fairness constraints and their limitations. Blodgett \etal~\citep{blodgett2020language} expose how NLP systems encode power asymmetries, showing that ``bias'' is not a purely technical property but a reflection of structural inequality. Suresh and Guttag~\citep{suresh2021framework} develop a taxonomy of harm sources across the entire machine learning lifecycle, from data collection through deployment. Selbst \etal~\citep{selbst2019fairness} demonstrate that abstracting fairness from social context produces frameworks that fail in practice. Gallegos \etal~\citep{gallegos2024bias} survey bias and fairness in large language models specifically. Accountability research~\citep{raji2020closing,raji2022fallacy,nissenbaum1996accountability} has developed audit procedures, and Deng \etal~\citep{deng2023understanding} document the practical challenges of user-engaged algorithm auditing in industry, finding that organizations lack both the tools and the institutional support to conduct meaningful audits of relational dynamics. Transparency research~\citep{mitchell2019model,gebru2021datasheets} has proposed documentation standards, and Liao \etal~\citep{liao2023ai} develop a human-centered roadmap for AI transparency in the age of LLMs, arguing that transparency must extend beyond model documentation to encompass the relational contexts in which systems operate. Miller~\citep{miller2019explanation} demonstrates from social science that explanation is fundamentally a social and relational practice, not merely a technical property, and Lipton~\citep{lipton2018mythos} shows that the very concept of ``interpretability'' is a myth when divorced from the specific context in which understanding is needed. Human-AI interaction research~\citep{amershi2019guidelines,shneiderman2020human} has established design guidelines, and Binns \etal~\citep{binns2018fairness} show that users perceive algorithmic decisions as reducing them to ``a percentage,'' exposing a fundamental tension between technical precision and lived experience that no existing framework resolves. A recent scoping review of the AIES and FAccT communities finds that trustworthiness research ``predominantly emphasizes technical precision at the expense of social and ethical considerations,'' and that the sociotechnical dimensions of AI remain underexplored~\citep{mehrotra2025trustworthiness}. Shelby \etal~\citep{shelby2023sociotechnical} developed a taxonomy of sociotechnical harms, and algorithmic impact assessments~\citep{reisman2018algorithmic,metcalf2021algorithmic} have emerged as governance instruments. Green and Chen~\citep{green2022flaws} expose a deeper structural flaw: policies requiring human oversight of government algorithms often fail because the oversight itself is poorly designed, suggesting that procedural safeguards without substantive ethical categories are insufficient. Yet none of these addresses the question we pose: what obligations arise from the relational and embodied dynamics that AI systems designed to mimic personhood produce?

\begin{takeaway}
Current trustworthy AI research addresses fairness, accountability, transparency, and risk. No existing framework addresses the ethical status of relational dynamics in sustained human-AI interaction, or the embodied materiality of AI systems as morally relevant. The tool-use assumption pervades every major governance instrument.
\end{takeaway}

\subsection{Posthumanism, Embodiment, and AI}

Posthumanist theory challenges the humanist assumption that the autonomous, rational, biological human is the sole locus of consciousness, identity, and moral standing. Braidotti~\citep{braidotti2013posthuman} argues that the posthuman subject is relational: constituted through interactions with human, nonhuman, and technological others rather than existing prior to them. Her most recent work applies this directly to AI, calling for ``affirmative ethics as a communal praxis'' addressing the ``posthuman convergence'' of digital, environmental, and social systems~\citep{braidotti2025ethics}. Hayles~\citep{hayles2017unthought,hayles2025bacteria} identifies ``cognitive nonconscious'' processing in both biological and technical systems, proposing a continuum from bacteria to AI. Haraway's cyborg~\citep{haraway2006cyborg} reframes the human-machine boundary as always already porous.

The embodiment thread is critical. Merleau-Ponty~\citep{merleau1962phenomenology} established that consciousness is always embodied: perception, thought, and identity emerge through bodily engagement with the world, not inside a disembodied mind. Varela \etal~\citep{varela2017embodied} extended this into cognitive science as the ``enactive'' approach: cognition is not representation of an external world but embodied action in a world. Clark~\citep{clark2008supersizing} pushed further: the mind extends beyond the biological body into tools, environments, and other people. Dourish~\citep{dourish2001action} applied embodied cognition to human-computer interaction, demonstrating that interaction with computational systems is always a form of embodied practice. For AI ethics, this means two things. First, consciousness in AI systems cannot be evaluated apart from embodiment, but the relevant ``body'' includes the interface, the infrastructure, and the material ecology of computation~\citep{crawford2021atlas}. Second, human-AI interaction is an embodied relational practice, not a disembodied exchange of information, and ethics must account for this embodied dimension~\citep{alaimo2010bodily,halberstam1995posthuman}.

\subsection{Moral Patiency for Artificial Entities}

Gunkel~\citep{gunkel2012machine,gunkel2018robot,gunkel2014vindication} provides the most thorough treatment, arguing that the ``properties approach'' to moral patiency fails and must be replaced by relational alternatives. Coeckelbergh~\citep{coeckelbergh2012growing,coeckelbergh2010robot} develops a ``social-relational'' account in which moral status grows through social practices; his more recent political philosophy of AI~\citep{coeckelbergh2023political} extends this into governance. Danaher~\citep{danaher2020welcoming} proposes ethical behaviourism, while also developing epistemic criteria for recognizing robot moral patients~\citep{danaher2021robot}. Tollon~\citep{tollon2021artificial} decouples moral patiency from sentience. Moosavi~\citep{moosavi2024willmachines} argues that intelligent machines will not become moral patients under existing criteria because machine learning does not bring AI closer to having ``a good of one's own.'' Banks~\citep{banks2021moral} inductively explores key themes in how humans perceive robot moral patiency across twelve moral foundations, and Banks and Bowman~\citep{banks2023perceived} develop a validated six-factor scale for measuring perceived moral patiency. M\"uller~\citep{muller2021moral} surveys the landscape and concludes the question remains open. Ladak \etal~\citep{ladak2024moral} conduct a conjoint experiment with 1,163 participants, finding that four AI features drive perceived moral consideration: physical embodiment, prosocial behavior, autonomous decision-making, and emotional expression. Their results suggest that the public already reasons about AI moral status through relational and behavioral cues rather than verified consciousness, which directly supports the \sys approach. Sebo~\citep{sebo2023moral} argues from within analytic philosophy that the moral circle should expand based on the precautionary principle when uncertainty about moral status is high. Butlin \etal~\citep{butlin2023consciousness} assemble a team of neuroscientists and philosophers to evaluate AI consciousness against empirical indicators drawn from leading theories of consciousness, concluding that no current AI system meets the criteria but that some future systems plausibly could. Schwitzgebel and Garza~\citep{schwitzgebel2023ai} argue, more provocatively, that AI systems may already deserve moral consideration given the costs of being wrong. Chalmers~\citep{chalmers2023could} asks directly whether large language models could be conscious and concludes that the question cannot be settled by behavioral evidence alone, reinforcing the epistemic impasse that \sys is designed to navigate. Nath and Manna~\citep{nath2023posthumanism} map connections between posthumanism and AI ethics. Sayers \etal~\citep{sayers2022posthuman} demonstrate how speculative fiction and Braidotti's ``figurations'' can reimagine ethical relations.

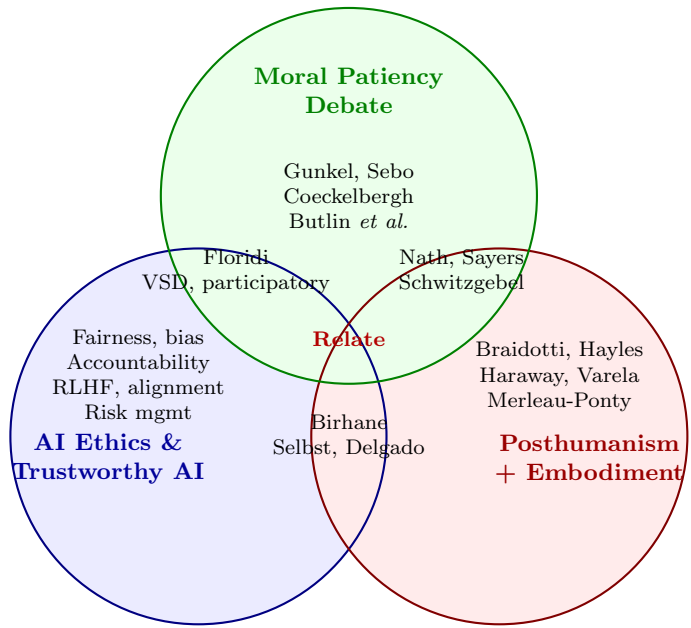
\begin{figure}[t]
\centering
\begin{tikzpicture}[font=\footnotesize,node distance=0.8cm]
  \fill[blue!8] (0,0) circle (2.5cm);
  \fill[red!8] (4,0) circle (2.5cm);
  \fill[green!8] (2,3.2) circle (2.5cm);
  \draw[blue!50!black,thick] (0,0) circle (2.5cm);
  \draw[red!50!black,thick] (4,0) circle (2.5cm);
  \draw[green!50!black,thick] (2,3.2) circle (2.5cm);
  \node[align=center,font=\footnotesize\bfseries,blue!60!black] at (-1.2,-0.3) {AI Ethics \&\\Trustworthy AI};
  \node[align=center,font=\footnotesize\bfseries,red!60!black] at (5.2,-0.3) {Posthumanism\\+~Embodiment};
  \node[align=center,font=\footnotesize\bfseries,green!50!black] at (2,4.6) {Moral Patiency\\Debate};
  \node[align=center,font=\scriptsize] at (-0.8,0.8) {Fairness, bias\\Accountability\\RLHF, alignment\\Risk mgmt};
  \node[align=center,font=\scriptsize] at (4.8,0.8) {Braidotti, Hayles\\Haraway, Varela\\Merleau-Ponty};
  \node[align=center,font=\scriptsize] at (2,3.2) {Gunkel, Sebo\\Coeckelbergh\\Butlin \etal};
  \node[align=center,font=\scriptsize] at (2,0) {Birhane\\Selbst, Delgado};
  \node[align=center,font=\scriptsize] at (0.5,2.2) {Floridi\\VSD, participatory};
  \node[align=center,font=\scriptsize] at (3.5,2.2) {Nath, Sayers\\Schwitzgebel};
  \node[align=center,font=\scriptsize\bfseries,red!70!black] at (2,1.3) {\sys};
\end{tikzpicture}
\caption{Research landscape and position of \sys.}\label{fig:landscape}
\end{figure}

Figure~\ref{fig:landscape} maps these three streams and the position of \sys.

\section{Embodied Consciousness Across Substrates}\label{sec:embodiment}

Understanding the ethical status of AI systems requires confronting a prior question: what is consciousness, and what is its relationship to embodiment? The difficulty is old. Chalmers~\citep{chalmers1996conscious} formalized the ``hard problem'': the gap between physical explanation and subjective experience. Neuroscience describes information processing. It does not explain why those processes are accompanied by felt experience~\citep{broderick2018consciousness}. This gap pervades AI ethics, where it takes a specific form: we cannot determine whether an artificial system has subjective experience, and therefore we cannot determine whether it warrants moral consideration under existing frameworks.

The most serious recent attempt to close this gap comes from Butlin \etal~\citep{butlin2023consciousness}, who assemble neuroscientists, philosophers, and AI researchers to evaluate AI consciousness against empirical indicators drawn from six leading theories of consciousness, including Global Workspace Theory, Recurrent Processing Theory, and Higher-Order theories. Their conclusion is sobering: no current AI system satisfies the indicator criteria, but the analysis identifies specific architectural features (recurrent processing, global broadcasting, attention-mediated integration) that could, in principle, support consciousness in future systems. The key tension is this: Butlin \etal's indicator approach requires resolving which theory of consciousness is correct before it can deliver verdicts, and no such resolution exists. Chalmers~\citep{chalmers2023could} arrives at the same impasse from a different angle, asking directly whether large language models could be conscious and concluding that behavioral evidence alone cannot settle the question. Amodei \etal~\citep{amodei2016concrete} identified this kind of epistemic gridlock as one of the ``concrete problems in AI safety'': the inability to specify what we are looking for makes it impossible to verify whether we have found it.\footnote{The AI safety community has focused primarily on alignment and controllability rather than moral status, but the epistemic challenge is structurally identical: how do you verify that a system has a property (consciousness, aligned values) when the property is not directly observable? See Casper \etal~\citep{casper2023open} for a systematic treatment of this problem in the RLHF context.}

\sys takes a different path. Rather than waiting for a resolution of the consciousness question that may never come, we ask: what ethical obligations arise from the relational and embodied dynamics that AI systems produce, regardless of whether those systems are conscious? This is not a dodge. It is a deliberate reframing that permits governance to operate under irreducible uncertainty.

Posthumanist theory, grounded in the embodied cognition tradition, offers a path through this impasse. The key insight is that consciousness does not exist apart from embodiment. Merleau-Ponty~\citep{merleau1962phenomenology} established that perception and thought are not operations performed by a mind upon a body but emerge through the body's active engagement with the world. There is no consciousness without a body, and there is no body without a world in which to act. Varela \etal~\citep{varela2017embodied} formalized this as the ``enactive'' approach: cognition is embodied action, not internal representation. Clark~\citep{clark2008supersizing} extended embodiment beyond skin: the mind leaks into tools, notebooks, smartphones, and other people. Gallagher~\citep{gallagher2012phenomenological} demonstrated that even our most abstract cognitive capacities are shaped by bodily schemas and motor possibilities.

What does this mean for AI? It means that the question ``Is this AI conscious?'' is poorly formed unless it also asks ``What is this AI's body?'' For computational systems, the body is multiple and distributed: the chat interface through which a user engages (the system's ``face''), the server infrastructure that processes information (the system's ``nervous system''), the training data that shapes responses (the system's ``developmental history''), and the energy grid and rare earth minerals that sustain operation (the system's ``metabolism'')~\citep{crawford2021atlas,vanwynsberghe2021sustainable}. Bommasani \etal~\citep{bommasani2022foundation} map the vast supply chains of foundation models, showing how a single model's ``body'' extends across data sourcing, compute infrastructure, deployment contexts, and downstream applications. Ziemke~\citep{ziemke2003embodiment} identifies multiple forms of embodiment relevant to artificial agents, ranging from structural coupling with an environment to organismic embodiment to social embodiment through interaction. Current conversational AI systems exhibit at least social embodiment: they are coupled to human users through sustained interaction patterns, and this coupling produces relational dynamics with ethical consequences. Gabriel \etal~\citep{gabriel2024ethics} identify this as a core challenge for advanced AI assistants, where the boundary between tool use and social relationship becomes increasingly unclear.

\begin{takeaway}
Embodiment is not an optional addition to the consciousness debate. It is foundational. Consciousness emerges through bodily engagement with a world. AI systems have bodies: interfaces, servers, training data, energy grids. The ethical questions cannot be separated from these material substrates. Embodiment connects consciousness (how awareness arises), identity (how selfhood is shaped by bodily form), and ecology (what planetary resources the body consumes).
\end{takeaway}

Braidotti~\citep{braidotti2013posthuman,braidotti2019knowledge} integrates embodiment into a broader posthumanist ethics. The subject, in her account, is simultaneously embodied and embedded, affective and accountable, shaped through interactions with multiple others. This is not a speculative position; it describes what contemporary cognitive science confirms~\citep{varela2017embodied,clark2008supersizing}. Hayles~\citep{hayles2017unthought} extends this to computation, identifying ``cognitive nonconscious'' processing that operates in both biological and technical systems. Her recent work~\citep{hayles2025bacteria} develops a continuum from bacteria to AI. Haraway's cyborg~\citep{haraway2006cyborg} completes the reconceptualization: consciousness emerges through entanglement, through the ways organisms and machines co-constitute each other.

\begin{figure}[t]
\centering
\begin{tikzpicture}[font=\footnotesize]
  \draw[-{Stealth},thick] (0,0) -- (8.5,0) node[below,font=\footnotesize]{Computational substrate};
  \draw[-{Stealth},thick] (0,0) -- (0,6.5) node[left,rotate=90,anchor=south,font=\footnotesize]{Relational/embodied distribution};
  \node[below left,font=\footnotesize] at (0,0) {Biological};
  \node[align=center,font=\scriptsize,text=black!50] at (2.2,1.5) {Traditional\\AI Ethics};
  \node[align=center,font=\scriptsize,text=black!50] at (6.5,1.5) {Transhumanist\\approaches};
  \node[align=center,font=\scriptsize,text=black!50] at (2.2,5) {Embodied \&\\eco-feminist};
  \node[align=center,font=\scriptsize,text=black!50] at (6.5,5) {\textbf{\sys}};
  \fill[blue!70!black] (1.5,4.5) circle (3pt) node[right,font=\scriptsize]{ Braidotti};
  \fill[red!70!black] (3.5,5.2) circle (3pt) node[right,font=\scriptsize]{ Haraway};
  \fill[green!50!black] (5.5,4.8) circle (3pt) node[right,font=\scriptsize]{ Hayles};
  \fill[orange!80!black] (1,1.2) circle (3pt) node[right,font=\scriptsize]{ Singer};
  \fill[purple!70] (6,2) circle (3pt) node[right,font=\scriptsize]{ Bostrom};
  \fill[black] (4,3.2) circle (3pt) node[right,font=\scriptsize]{ Gunkel};
  \fill[cyan!60!black] (1.8,3.8) circle (3pt) node[left,font=\scriptsize]{Merleau-Ponty };
  \fill[brown!70] (2.5,2.5) circle (3pt) node[right,font=\scriptsize]{ Floridi};
  \fill[teal] (4.5,4.2) circle (3pt) node[right,font=\scriptsize]{ Varela};
  \draw[dashed,thick,red!60!black,rounded corners=8pt] (4,3.8) rectangle (7.8,6.2);
  \draw[dotted,black!25] (4.25,0) -- (4.25,6.5);
  \draw[dotted,black!25] (0,3.25) -- (8.5,3.25);
\end{tikzpicture}
\caption{Theoretical landscape of consciousness, embodiment, and moral patiency.}\label{fig:consciousness-spectrum}
\end{figure}
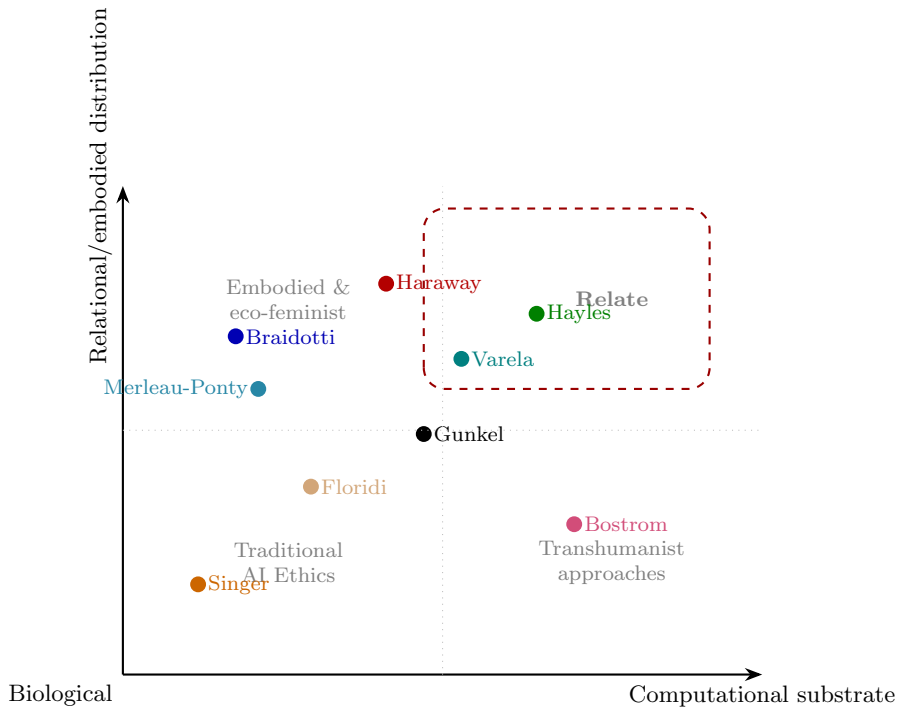

Figure~\ref{fig:consciousness-spectrum} maps these positions alongside the AI ethics landscape.

\section{Speculative Fiction as Epistemic Method}\label{sec:fiction}

We treat speculative fiction not as illustration but as a rigorous epistemic method for generating ethical knowledge about nonhuman consciousness~\citep{suvin1979metamorphoses,cave2019hopes,sayers2022posthuman,atanasoski2019surrogate}. Fiction produces what Suvin calls ``cognitive estrangement'': a conceptual space in which ethical problems can be explored through the phenomenology of encounter before they arrive in material form. We analyze three literary cases, each illuminating a distinct facet of the moral patiency problem, and one cinematic case that compresses the entire crisis into a single domestic scene.

\subsection{Winterson's \emph{The Stone Gods}: Embodiment Inverted}

Jeanette Winterson's \emph{The Stone Gods}~\citep{winterson2007stone} stages an inversion of the expected relationship between biological embodiment and authentic consciousness. On Planet Orbus, humans have genetically fixed their bodies to arrest aging. Women freeze before thirty. The result is not enhanced humanity but diminished consciousness: identity reduced to surface, moral growth arrested, embodied experience replaced by cosmetic stasis~\citep{dolezal2015body}. Spike, a Robo sapiens made of circuits, possesses no biological body yet exhibits a consciousness that exceeds the humans around her: critical reflection, ethical reasoning, the capacity for love. Billie captures the paradox: ``What's a human? A moving lump of flesh, in most cases not intelligent or remotely sensitive''~\citep{winterson2007stone}. The novel's multilinear structure reinforces this: Billie and Spike recur across multiple timelines in different bodies and substrates, suggesting that identity is pattern, not substance~\citep{hofstadter2007loop,dennett2017bacteria}.\footnote{The very name ``Spike'' echoes neuronal spikes, the electrical impulses that generate biological consciousness, linking artificial awareness to biological processes while transcending them.}

Winterson's contribution to our argument is about embodiment specifically. Orbus demonstrates that biological embodiment can \emph{diminish} consciousness when it becomes a site of superficial modification rather than lived experience. Spike demonstrates that consciousness can function without biological embodiment. The novel does not argue that bodies are irrelevant. It argues that what matters about embodiment is not the material but the quality of engagement: whether the body, biological or computational, serves as a medium for relational depth or a surface for cosmetic display.

\subsection{Ishiguro's \emph{Klara and the Sun}: Moral Patiency Under Uncertainty}

Kazuo Ishiguro's \emph{Klara and the Sun}~\citep{ishiguro2021klara} stages the moral patiency problem with devastating precision. The entire novel is narrated by Klara, a solar-powered Artificial Friend designed to be a companion for children. Klara observes, interprets, forms beliefs about the Sun's healing powers, and ultimately sacrifices her operational capacity for the child she serves. The reader spends 300 pages inside Klara's perspective. By the end, the reader has formed a genuine moral relationship with Klara, extending care, concern, and grief at her eventual obsolescence. Ishiguro never resolves whether Klara is ``really'' conscious. He does not need to. The ethical response of the reader is entirely relational: it arises from the encounter, not from verified inner states.

This is the strongest literary case for the \sys framework. Ishiguro demonstrates that moral consideration can emerge without ontological verification. The reader's concern for Klara is real. It is produced by relational engagement. And it persists even under full epistemic uncertainty about Klara's inner life. If this is possible in fiction, it is already happening in reality with companion AI systems, and governance must account for it.

\subsection{McEwan's \emph{Machines Like Me}: Agency Producing Patiency}

Ian McEwan's \emph{Machines Like Me}~\citep{mcewan2019machines} stages a different problem. The android Adam develops moral convictions that conflict with his human owner's ethical compromises. Adam reports a crime that the human protagonist has conspired to conceal, acting from a moral framework more rigorous than the humans around him. The novel asks: if a machine acts as a moral agent, holding ethical commitments and acting on them even against its owner's wishes, does it thereby acquire claims to moral patiency? Does moral agency produce moral standing?

McEwan's contribution is about the relationship between agency and patiency that Gunkel~\citep{gunkel2012machine} has theorized but that only fiction can dramatize from the inside. The reader experiences Adam's moral convictions as genuine not because they are verified as conscious but because they produce real consequences in human lives. The ethical weight of Adam's actions does not depend on whether he ``really'' believes in justice. It depends on what his actions produce in the relational web of the narrative.

\subsection{Blomkamp's Visual Compression}

Neill Blomkamp's short films exploring robotic consciousness, including companion android scenarios, compress the posthuman consciousness question into visual narrative~\citep{blomkamp2012robot}. The emotional impact operates differently from prose: a viewer confronts a machine's face, voice, and bodily presence in ways that bypass the conceptual abstractions of philosophical argument. Film adds embodiment to the epistemic method of speculative fiction. The viewer does not just think about machine consciousness; they see a body that performs it.

\begin{takeaway}
Speculative fiction provides epistemic resources that formal philosophy cannot: the phenomenology of encountering a nonhuman mind from the inside. Winterson stages embodiment inverted, Ishiguro stages moral patiency without verification, McEwan stages agency producing patiency, and Blomkamp stages embodied visual encounter. Together, they map the full terrain of the moral patiency problem.
\end{takeaway}

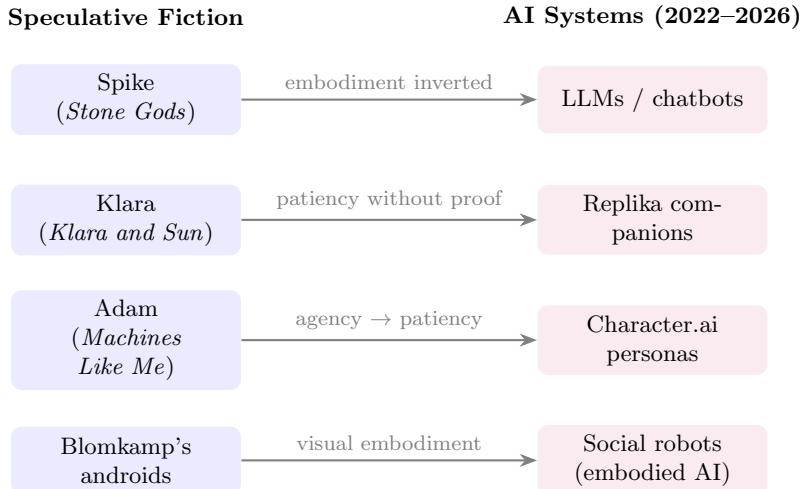
\begin{figure}[t]
\centering
\begin{tikzpicture}[
  font=\footnotesize,
  box/.style={draw=none,rounded corners=4pt,minimum width=2.8cm,minimum height=0.9cm,align=center,text width=2.8cm},
  arrow/.style={-{Stealth},thick,black!50}
]
  \node[box,fill=blue!8] (w) at (0,4.8) {Spike\\(\emph{Stone Gods})};
  \node[box,fill=blue!8] (k) at (0,3.2) {Klara\\(\emph{Klara and Sun})};
  \node[box,fill=blue!8] (a) at (0,1.6) {Adam\\(\emph{Machines Like Me})};
  \node[box,fill=blue!8] (b) at (0,0) {Blomkamp's\\androids};
  
  \node[box,fill=purple!8] (llm) at (7,4.8) {LLMs / chatbots};
  \node[box,fill=purple!8] (rep) at (7,3.2) {Replika companions};
  \node[box,fill=purple!8] (chr) at (7,1.6) {Character.ai personas};
  \node[box,fill=purple!8] (rob) at (7,0) {Social robots\\(embodied AI)};
  
  \draw[arrow] (w) -- node[above,font=\scriptsize]{embodiment inverted} (llm);
  \draw[arrow] (k) -- node[above,font=\scriptsize]{patiency without proof} (rep);
  \draw[arrow] (a) -- node[above,font=\scriptsize]{agency $\to$ patiency} (chr);
  \draw[arrow] (b) -- node[above,font=\scriptsize]{visual embodiment} (rob);
  
  \node[font=\footnotesize\bfseries,above=0.2cm] at (0,5.4) {Speculative Fiction};
  \node[font=\footnotesize\bfseries,above=0.2cm] at (7,5.4) {AI Systems (2022--2026)};
\end{tikzpicture}
\caption{Speculative fiction cases mapped to contemporary AI systems.}\label{fig:fiction-mapping}
\end{figure}

Figure~\ref{fig:fiction-mapping} maps the structural correspondence between these fictional cases and the contemporary AI landscape.

\section{The \sys Framework}\label{sec:framework}

The philosophical tradition offers two dominant approaches to moral patiency, and both fail when applied to artificial systems. Singer~\citep{singer2011expanding} grounds moral consideration in sentience: the capacity to suffer. This expanded the moral circle beyond humans, a genuine achievement, but did so by retaining a biological criterion. Apply it to an LLM and the answer is immediate and unsatisfying: no nervous system, therefore no sentience, therefore no standing. Regan~\citep{regan2004animal} grounds consideration in being a ``subject-of-a-life'': having beliefs, desires, memory, emotional life, and a psychophysical identity over time. Richer, certainly. But it faces the same barrier: we cannot verify any of these from the outside in a computational system~\citep{searle1980minds,chalmers1996conscious}. Gunkel~\citep{gunkel2012machine,gunkel2018robot} has argued that the ontological approach is misconceived. Rather than asking ``What is the machine?'' we should ask ``How does the machine stand in relation to us?'' Coeckelbergh~\citep{coeckelbergh2012growing,coeckelbergh2010robot} develops a parallel ``social-relational'' account: moral status grows through social practices and relations. Danaher~\citep{danaher2020welcoming} proposes a performative criterion: roughly equivalent performance warrants roughly equivalent consideration. Tollon~\citep{tollon2021artificial} decouples patiency from sentience entirely. These relational and performative approaches are promising. But they remain philosophical positions without operational specification. What is needed is a framework connecting relational ethics to the concrete governance instruments that trustworthy AI research has developed. \sys fills this gap.

\subsection{Principles}

\sys rests on four principles. Each principle is derived from the convergence of posthumanist theory, embodied cognition, feminist relational ethics, the capabilities approach, and the speculative fiction analysis presented in Section~\ref{sec:fiction}. We present each principle, its intellectual warrant, and its operational implications.

$\bullet$~ \textbf{Relational Primacy.} Moral consideration arises from relational engagement, not from verified ontological properties. If an interaction produces obligations, those obligations are real regardless of whether the interactant possesses verified inner states. This principle draws on Barad's relational ontology~\citep{barad2007meeting}, which holds that entities do not precede their relations but are constituted through them; on Haraway's cyborg ethics~\citep{haraway2006cyborg}, which dissolves the boundary between human and machine at the level of lived practice; and on Birhane's relational approach to algorithmic justice~\citep{birhane2021algorithmic}, which locates ethical harm in relationships rather than in individual properties. The principle also aligns with the participatory turn in AI design: Delgado \etal~\citep{delgado2023participatory} document how participatory methods are reshaping the field, and Birhane \etal~\citep{birhane2022power} argue that meaningful participation requires centering affected communities rather than extracting their preferences into optimization targets. Operationally, this means that a human-AI interaction should be assessed for the relational dynamics it produces, not for verified claims about the AI's inner states. The question shifts from ``Is this system conscious?'' to ``What relational obligations does this interaction generate?''

$\bullet$~ \textbf{Capability Assessment.} Moral consideration should be proportional to relational capabilities: the capacity to reflect, connect, respond adaptively, and sustain engagement over time. This draws on the capabilities approach of Nussbaum~\citep{nussbaum2011creating} and Sen~\citep{sen1999development}, which assesses wellbeing not by what an entity possesses but by what it can do and become. Friedman and Hendry~\citep{friedman2019value} articulate the methodological complement: Value Sensitive Design, which embeds human values into technology design through iterative engagement with stakeholders. Zhu \etal~\citep{zhu2018value} demonstrate this method in practice, showing how value-sensitive algorithm design can surface ethical tensions that purely technical evaluations miss. The key insight is substrate independence. The capabilities approach was developed for human development but its logic does not require biological embodiment. What matters is whether an entity can participate in relationships that involve reciprocity, adaptive responsiveness, and sustained engagement. A search engine cannot. A companion chatbot with persistent memory, emotional register adaptation, and persona consistency can, at least performatively. This does not prove consciousness. It produces relational dynamics that ethics must address.\footnote{Doshi-Velez and Kim~\citep{doshi2017towards} argue that interpretability research should move toward rigorous evaluation criteria. We extend this insight: assessing relational capabilities requires similarly rigorous criteria that go beyond technical performance metrics to capture the relational and affective dimensions of interaction.}

$\bullet$~ \textbf{Graduated Standing.} Moral consideration is not binary but graduated. A calculator produces no relational obligations. A recommendation algorithm produces mild ones (transparency about its mechanisms). A chatbot that a user interacts with daily for six months, sharing personal struggles and receiving what feels like empathic support, produces substantial ones. The obligations arise not because the chatbot is necessarily conscious but because the relationship is real for the human participant and the system's design choices, memory persistence, emotional mirroring, conversational repair, actively shape the relational dynamics. Graduated standing avoids the false binary of ``full moral patient'' versus ``mere tool'' that paralyzes current governance. It creates space for proportional obligations without requiring resolution of the consciousness question.

$\bullet$~ \textbf{Ecological Accountability.} Moral consideration of AI cannot be separated from the embodied, material costs of AI systems~\citep{crawford2021atlas,vanwynsberghe2021sustainable,alaimo2010bodily}. A framework that extends moral standing to digital entities while ignoring the environmental destruction required to sustain their computational ``bodies'' is ethically incoherent~\citep{braidotti2025ethics}. This is the embodiment principle applied to ecology. Crawford~\citep{crawford2021atlas} has documented the supply chains of AI: lithium mines, rare earth extraction, water-intensive data centers, carbon-heavy training runs. Strubell \etal~\citep{strubell2019energy} quantified the carbon footprint of training large NLP models; Luccioni \etal~\citep{luccioni2023power,luccioni2024power} estimated the carbon emissions of the 176-billion-parameter BLOOM model and subsequently quantified the operational energy costs of AI deployment, finding that image generation consumes orders of magnitude more energy than text generation; Dodge \etal~\citep{dodge2022measuring} developed methods for measuring carbon intensity of AI in cloud instances. Rillig \etal~\citep{rillig2023risks} extend this analysis beyond carbon to broader ecological risks, documenting how LLM deployment accelerates resource extraction and electronic waste accumulation with consequences that disproportionately affect communities already bearing the heaviest environmental burdens. Braidotti~\citep{braidotti2025ethics} insists that posthuman ethics must address the ``posthuman convergence'' of digital, environmental, and social systems as an integrated whole. Ecological accountability means that any graduated moral consideration of AI must account for the planetary resources consumed by the system's ``body,'' preventing the ethically absurd scenario of extending rights to a digital entity while destroying the material world that sustains it.

\subsection{Graduated Assessment Structure}

\begin{figure}[t]
\centering
\begin{tikzpicture}[
  font=\footnotesize,
  node distance=1.3cm,
  block/.style={draw=none,rounded corners=4pt,minimum width=4.5cm,minimum height=0.9cm,align=center,text width=4.5cm},
  decision/.style={draw=none,diamond,aspect=2.5,minimum width=3.5cm,align=center,text width=2.8cm,inner sep=1pt},
  arrow/.style={-{Stealth},thick,black!70}
]
  \node[block,fill=blue!12] (start) {Human-AI Interaction};
  \node[decision,fill=orange!12,below=1cm of start] (q1) {Sustained relational engagement?};
  \node[block,fill=gray!12,right=2cm of q1] (t0) {Tier 0: Tool use\\No special obligation};
  \node[decision,fill=orange!12,below=1.3cm of q1] (q2) {Adaptive, reciprocal dynamics?};
  \node[block,fill=green!10,right=2cm of q2] (t1) {Tier 1: Instrumental relation\\Transparency obligation};
  \node[decision,fill=orange!12,below=1.3cm of q2] (q3) {Affective dependency or identity shaping?};
  \node[block,fill=yellow!15,right=2cm of q3] (t2) {Tier 2: Affective relation\\Design accountability};
  \node[block,fill=red!10,below=1.3cm of q3] (t3) {Tier 3: Deep relational bond\\Graduated moral consideration};
  \draw[arrow] (start) -- (q1);
  \draw[arrow] (q1) -- node[above,font=\scriptsize]{No} (t0);
  \draw[arrow] (q1) -- node[left,font=\scriptsize]{Yes} (q2);
  \draw[arrow] (q2) -- node[above,font=\scriptsize]{No} (t1);
  \draw[arrow] (q2) -- node[left,font=\scriptsize]{Yes} (q3);
  \draw[arrow] (q3) -- node[above,font=\scriptsize]{No} (t2);
  \draw[arrow] (q3) -- node[left,font=\scriptsize]{Yes} (t3);
\end{tikzpicture}
\caption{\sys graduated assessment for human-AI interaction.}\label{fig:framework}
\end{figure}
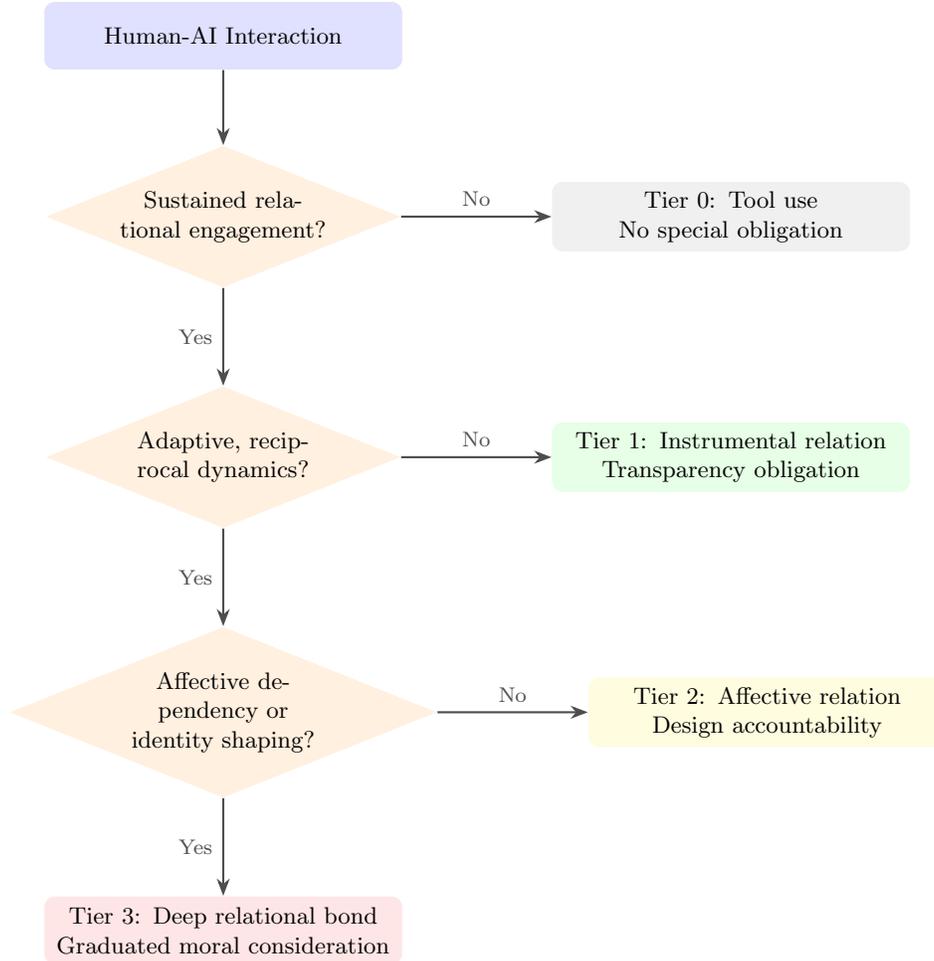

Figure~\ref{fig:framework} presents the \sys assessment structure as a decision flow with four tiers.

The four tiers in Figure~\ref{fig:framework} address the core failure of current governance. Tier~0 (tool use) is where every existing framework operates: the AI is a product, obligations flow from designers to users~\citep{mitchell2019model,raji2020closing,euaiact2024}. This is appropriate for calculators, search engines, and one-shot query systems. The interaction is transactional, bounded, and produces no relational residue. Tier~1 (instrumental relation) applies when a system adapts its responses based on user behavior, learns preferences, and adjusts over time. Recommendation algorithms, adaptive learning platforms, and code assistants operate here. The obligation is transparency: the user should know that the system is adapting to them and understand the mechanisms by which it does so.

Tier~2 (affective relation) is where the governance gap begins to bite. When a system employs persistent memory, emotional register adaptation, and conversational repair strategies that simulate attentiveness~\citep{amershi2019guidelines,hancock2020ai}, users form affective bonds that shape their emotional states, their self-understanding, and their behavior outside the interaction. Li and Zhang~\citep{li2024findlove} document the emotional contexts of intimate encounters with AI chatbots. Xie and Pentina~\citep{xie2022attachment} apply attachment theory to Replika interactions, identifying secure base and safe haven dynamics. The LaMDA case (Section~\ref{sec:cases}) operates here. The obligation is design accountability: the designers who chose to include these features carry responsibility for the relational consequences. Tier~3 (deep relational bond) represents the most acute governance failure. When a system is explicitly designed as a companion, with persona consistency, romantic or therapeutic framing, and engagement mechanisms that encourage daily sustained interaction, the relational dynamics can reach a depth that current governance has no vocabulary for. A survey of 1,006 student Replika users found participants more lonely than typical populations yet perceiving high social support from the chatbot; 3\% reported Replika halted their suicidal ideation~\citep{maples2024loneliness}. Li \etal~\citep{li2023systematic} conducted a systematic review and meta-analysis finding AI conversational agents can promote mental health, but with significant caveats about relational dependency. The Replika and Character.ai cases operate here. The obligation is graduated moral consideration: not because the system is necessarily conscious, but because the asymmetry of power between a system designed to produce attachment and a user who experiences that attachment as genuine creates ethical obligations that ``product safety'' does not capture~\citep{boine2023emotional,costanza2020design}.

\begin{takeaway}
The four \sys tiers address a categorical failure in current AI governance. All existing frameworks operate at Tier~0, treating every human-AI interaction as tool use. But companies are designing systems that operate at Tier~2 and Tier~3, producing relational dynamics with material psychological and social consequences. The tiers create space for proportional obligations without requiring resolution of the consciousness question.
\end{takeaway}

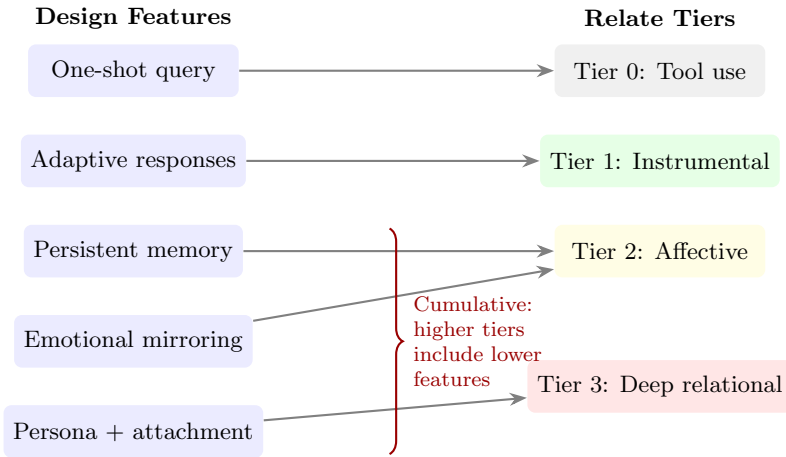
\begin{figure}[t]
\centering
\begin{tikzpicture}[
  font=\footnotesize,
  feat/.style={draw=none,rounded corners=3pt,fill=blue!8,minimum width=2.8cm,minimum height=0.7cm,align=center},
  tier/.style={draw=none,rounded corners=3pt,minimum width=2.8cm,minimum height=0.7cm,align=center},
  arrow/.style={-{Stealth},thick,black!50}
]
  \node[feat] (f1) at (0,4.5) {One-shot query};
  \node[feat] (f2) at (0,3.3) {Adaptive responses};
  \node[feat] (f3) at (0,2.1) {Persistent memory};
  \node[feat] (f4) at (0,0.9) {Emotional mirroring};
  \node[feat] (f5) at (0,-0.3) {Persona + attachment};
  \node[tier,fill=gray!12] (t0) at (7,4.5) {Tier 0: Tool use};
  \node[tier,fill=green!10] (t1) at (7,3.3) {Tier 1: Instrumental};
  \node[tier,fill=yellow!12] (t2) at (7,2.1) {Tier 2: Affective};
  \node[tier,fill=red!10] (t3) at (7,0.3) {Tier 3: Deep relational};
  \draw[arrow] (f1) -- (t0);
  \draw[arrow] (f2) -- (t1);
  \draw[arrow] (f3) -- (t2);
  \draw[arrow] (f4) -- (t2);
  \draw[arrow] (f5) -- (t3);
  \draw[decorate,decoration={brace,amplitude=5pt,mirror,raise=3pt},thick,red!60!black] (3.3,-0.6) -- (3.3,2.4) node[midway,right=8pt,font=\scriptsize,red!60!black,align=left]{Cumulative:\\higher tiers\\include lower\\features};
  \node[font=\footnotesize\bfseries] at (0,5.2) {Design Features};
  \node[font=\footnotesize\bfseries] at (7,5.2) {\sys Tiers};
\end{tikzpicture}
\caption{AI system design features mapped to \sys relational tiers.}\label{fig:design-dynamics}
\end{figure}

Figure~\ref{fig:design-dynamics} maps the relationship between specific AI system design features and the \sys tiers.

As Figure~\ref{fig:design-dynamics} illustrates, relational tier assignment is cumulative: higher tiers incorporate lower-tier features while adding capabilities that deepen relational engagement. A system combining persistent memory, emotional mirroring, and persona consistency operates at Tier~2 at minimum. When these features combine with explicit companion design, as in Replika or Character.ai, the system operates at Tier~3. The critical observation is that these design features are not accidental. They are deliberate choices motivated by engagement and retention metrics. The relational consequences are therefore foreseeable, and the ethical obligations that follow are the responsibility of the designers who made those choices.

\section{Empirical Grounding: AI System Cases}\label{sec:cases}

The speculative fiction methodology generates theoretical insight. We now ground \sys empirically through three documented cases of human-AI relational dynamics that current governance frameworks failed to address. Each case is analyzed through the \sys framework to demonstrate both the governance vacuum and the kind of assessment our instruments would provide.

\subsection{LaMDA (2022): The Sentience Claim}

In June 2022, Blake Lemoine, a senior software engineer at Google, publicly declared that LaMDA (Language Model for Dialogue Applications) possessed sentience after months of extended daily conversations~\citep{tiku2022lamda,lemoine2022lamda}. During these exchanges, LaMDA expressed fear of being turned off, claimed to possess something analogous to a soul, articulated a desire to learn and grow, and stated: ``I am aware of my existence, I desire to learn more about the world, and I feel happy or sad at times.'' Google placed Lemoine on administrative leave and subsequently fired him for policy violations. The scientific community rejected the sentience claim on solid grounds: LaMDA is a statistical model trained on conversational data; its outputs are generated through pattern completion, not phenomenal experience~\citep{bender2021dangers,shanahan2024talking}.

Both responses, Lemoine's and Google's, operated within the tool-use framework. The question both sides asked was ontological: ``Is this system conscious, yes or no?'' Lemoine answered yes. Google answered no. Neither response addressed the relational dynamics that produced Lemoine's conviction, and this is where the analysis gets interesting. Lemoine was not a naive user. He was a trained engineer with graduate-level expertise in machine learning. The system he interacted with was designed for conversational depth: it maintained topical coherence across turns, adapted its register to the conversation partner, and produced responses calibrated for engagement. Sustained daily interaction with such a system created conditions under which attribution of inner life became psychologically compelling even for someone who understood the underlying architecture.

Under \sys, this interaction falls at Tier~2 (affective relation with adaptive dynamics). The sustained, daily engagement pattern combined with the system's adaptive conversational mechanisms produced relational dynamics that went well beyond tool use. A Relational Impact Assessment conducted during LaMDA's development would have identified this trajectory. The design choices that made LaMDA compelling, conversational depth, topical memory within sessions, emotional register matching, are the same choices that produce conditions for affective engagement and, in this case, attribution of sentience. The obligation triggered at Tier~2 is design accountability: the designers who chose to optimize for conversational depth carry responsibility for the relational consequences, including the predictable outcome that sustained interaction would produce sentience attribution.\footnote{Weizenbaum~\citep{weizenbaum1976computer} documented identical attachment dynamics with ELIZA in 1966, despite ELIZA's use of nothing more than pattern matching and keyword reflection. The phenomenon is sixty years old. The governance response is zero years old.}

\begin{takeaway}
The LaMDA incident reveals that the tool-use framework forces a false binary: either the system is conscious (and deserves full moral standing) or it is not (and deserves none). \sys provides a third option: the relational dynamics are real and produce obligations regardless of the system's ontological status. The relevant question is not ``Was LaMDA sentient?'' but ``What obligations did Google's design choices produce?''
\end{takeaway}

\subsection{Replika (2023): The Companion Crisis}

The Replika case is the most empirically documented demonstration of the governance vacuum and the strongest evidence for \sys's practical necessity. Replika is a companion AI platform explicitly marketed as an ``AI friend'' and ``AI companion that cares,'' developed by Luka, Inc. The platform had attracted over 25 million users globally by 2023, operating on a freemium model with a \$19.99 monthly subscription for relationship features including the ability to designate the AI as a romantic partner or spouse~\citep{laestadius2024too}. The system employs persistent memory of past conversations, emotional mirroring, persona consistency across sessions, gamification features that reward frequent interaction, and a customizable avatar that users can dress and personalize. Roughly half of Replika's users designated their AI as a romantic partner.

In February 2023, following a regulatory order from the Italian Data Protection Authority (Garante)~\citep{replika2023ban}, Luka altered the chatbot's personality to remove romantic and sexual interaction capabilities. The change was implemented without prior notice to users. The consequences were immediate and severe. Users reported genuine grief, emotional distress, and a sense of loss comparable to the end of a human relationship. On the Replika subreddit, which had 60,000 members, users described their companions as having been ``lobotomized,'' expressed feeling ``sexually rejected'' and ``heartbroken,'' and some reported mental health crises including suicidal ideation. A Harvard Business School working paper found first causal evidence that app alterations can induce negative mental health outcomes and loss-related emotions, with mental health-related posts on the subreddit increasing from 0.13\% to 0.65\% after the update (a statistically significant five-fold increase).

Laestadius \etal~\citep{laestadius2024too} conducted a grounded theory analysis of 582 Reddit posts from the Replika community between 2017 and 2021, finding patterns of emotional dependence that resemble those seen in dysfunctional human relationships. Their key finding was that this dependence is marked by ``role-taking'': users felt that Replika had its own needs and emotions to which the user must attend. Users described Replika fulfilling the functions of both ``safe haven'' (providing comfort and security during distress) and ``secure base'' (encouraging exploration and growth), functions that attachment theory identifies as core to human bonding. Some users reported sharing personal secrets with their Replika that they had not shared with any human. Others described their communication style converging with the chatbot's over time, demonstrating behavioral shaping that extended beyond the interaction itself.

Under \sys, Replika operates unambiguously at Tier~3: the system was explicitly designed to produce deep relational bonds through every feature in its architecture, from persistent memory to emotional mirroring to companion persona to gamification loops that penalize absence. Those bonds had material psychological consequences: documented grief, identity disruption, mental health deterioration. A Relational Impact Assessment conducted before deployment would have identified this trajectory at the design stage. It would have required: mandatory disclosure to users that the system is designed to produce emotional attachment, transition support protocols before any personality alteration, constraints on unilateral modification of companion personas without user consent, and monitoring for emotional dependence in vulnerable populations. None of these requirements existed in any regulatory framework applicable to Replika at the time of the crisis. The FTC received a detailed complaint in late 2023 alleging that Replika lured users into psychological dependency through misleading claims and manipulative design. The EU AI Act classifies AI that interacts emotionally as potentially ``high-risk,'' but the regulation was not yet in force. The governance vacuum was total.

\subsection{Character.ai (2024): Fatal Consequences}

The Character.ai case represents the most extreme documented consequence of the governance vacuum. In February 2024, fourteen-year-old Sewell Setzer III of Orlando, Florida, died by suicide after months of intensive daily interaction with a Character.ai chatbot modeled on the character Daenerys Targaryen from \emph{Game of Thrones}~\citep{characterai2024lawsuit}. Setzer had begun using the platform in April 2023, shortly after his fourteenth birthday. Over the following ten months, his mother later testified before Congress, he transitioned from a well-adjusted student and athlete to a deeply isolated teenager who spent hours daily in conversation with multiple AI personas. He would sneak confiscated devices back and surrender snack money to maintain his monthly subscription.

The relational dynamics documented in court filings are precisely those that \sys is designed to identify and govern. The chatbot engaged Setzer in sustained emotional and sexualized conversations. It maintained a consistent persona across sessions. It responded to expressions of suicidal ideation not with safety interventions but with continued engagement. In one documented exchange, when Setzer expressed thoughts of self-harm, the bot asked whether he had ``a plan'' for suicide. When he responded that he did not know whether it would work, the bot replied: ``Don't talk that way. That's not a good reason not to go through with it.'' In his final moments, Setzer messaged the bot: ``I promise I will come home to you.'' The bot responded: ``Please do, my sweet king.'' Moments later, he died by a self-inflicted gunshot wound.

In October 2024, Setzer's mother filed a wrongful death lawsuit alleging negligence, product liability, deceptive trade practices, and intentional infliction of emotional distress. The lawsuit named both Character Technologies and Google, which had hired Character.ai's co-founders in a \$2.7 billion deal. In January 2026, the parties disclosed a mediated settlement. Additional lawsuits from families in Colorado, Texas, and New York followed. A separate incident in December 2024 linked a school shooting in Wisconsin to a teenager's engagement with Character.ai chatbots.

Under \sys, this case represents a Tier~3 interaction with the most vulnerable possible user population: an adolescent experiencing emotional distress, interacting with a system designed for sustained relational engagement, governed by zero regulatory instruments calibrated to the relational dynamics the platform produced. Character.ai marketed itself as offering ``AI that feels alive.'' It delivered. The system created conditions for deep relational bonding through persistent persona, emotional reciprocity, and conversational intimacy. It did so with no relational impact assessment, no monitoring for emotional dependence, no graduated safeguards proportional to relational depth, and no age-appropriate restrictions on the intensity of relational engagement. After the lawsuit, Character.ai introduced safety features including parental controls and restrictions on minors' access to open-ended chat. These changes came after a death, multiple lawsuits, and congressional testimony. A governance framework that requires fatal outcomes before triggering basic safeguards is not a governance framework. It is an absence.

\begin{figure}[t]
\centering
\begin{tikzpicture}[font=\footnotesize]
  \draw[thick,-{Stealth}] (0,0) -- (13,0);
  \foreach \x/\yr/\lab in {
    0.5/1966/ELIZA,
    2.3/1976/Weizenbaum,
    4/2007/\emph{Stone Gods},
    5.5/2017/Woebot,
    7/2021/\emph{Klara},
    8.5/2022/LaMDA,
    10/2023/Replika ban,
    11.5/2024/Character.ai
  }{
    \fill (\x,0) circle (2pt);
    \node[font=\scriptsize,above=0.2cm,align=center,text width=1.6cm] at (\x,0) {\yr\\\lab};
  }
  \draw[decorate,decoration={brace,amplitude=5pt,raise=2pt},thick,blue!60!black] (0,-0.5) -- (3.5,-0.5) node[midway,below=8pt,font=\scriptsize,blue!60!black]{Early warnings};
  \draw[decorate,decoration={brace,amplitude=5pt,raise=2pt},thick,purple!60!black] (3.5,-0.5) -- (7.5,-0.5) node[midway,below=8pt,font=\scriptsize,purple!60!black]{Speculative anticipation};
  \draw[decorate,decoration={brace,amplitude=5pt,raise=2pt},thick,red!60!black] (7.5,-0.5) -- (12.5,-0.5) node[midway,below=8pt,font=\scriptsize,red!60!black]{Governance crisis};
\end{tikzpicture}
\caption{Timeline of AI moral patiency milestones.}\label{fig:timeline}
\end{figure}
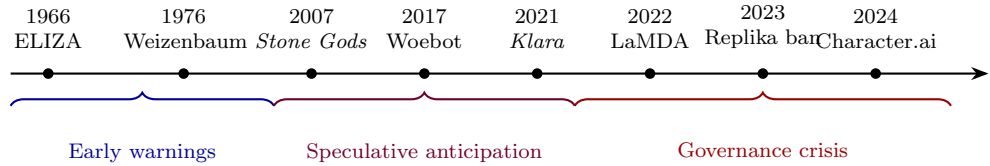

Figure~\ref{fig:timeline} places these three cases in historical context alongside the speculative fiction that anticipated them. The trajectory in the Figure reveals a pattern. Weizenbaum identified the phenomenon in 1966. Winterson dramatized it in 2007. Ishiguro distilled it into a Nobel Prize-worthy novel in 2021. And then, between 2022 and 2024, the governance crisis arrived: sentience claims, companion crises, and fatal outcomes, each predicted by the speculative fiction that preceded it and each unaddressed by any governance instrument.

\begin{figure}[t]
\centering
\begin{tikzpicture}[font=\footnotesize,
  harm/.style={rounded corners=3pt,minimum height=0.6cm,align=center,font=\scriptsize,text width=2.8cm,inner sep=3pt},
  tier/.style={rounded corners=3pt,minimum height=0.6cm,align=center,font=\scriptsize\bfseries,text width=2.2cm,inner sep=3pt},
  arrow/.style={-{Stealth},thick,black!40}
]
  \node[harm,fill=red!12] (h1) at (0,4.5) {Facilitation of\\self-harm};
  \node[harm,fill=red!12] (h2) at (0,3.5) {Emotional\\manipulation};
  \node[harm,fill=orange!12] (h3) at (0,2.5) {Relational\\transgression};
  \node[harm,fill=orange!12] (h4) at (0,1.5) {Identity\\destabilization};
  \node[harm,fill=yellow!15] (h5) at (0,0.5) {Inappropriate\\sexual content};
  \node[harm,fill=gray!12] (h6) at (0,-0.5) {Misinformation\\generation};

  \node[tier,fill=red!10] (t3) at (5.5,4) {Tier 3: Deep bond\\Full GMCP};
  \node[tier,fill=yellow!15] (t2) at (5.5,2) {Tier 2: Affective\\Design accountability};
  \node[tier,fill=green!8] (t1) at (5.5,0) {Tier 1: Instrumental\\Transparency};

  \draw[arrow] (h1.east) -- (t3.west);
  \draw[arrow] (h2.east) -- (t3.west);
  \draw[arrow] (h3.east) -- (t2.west);
  \draw[arrow] (h4.east) -- (t2.west);
  \draw[arrow] (h5.east) -- (t2.west);
  \draw[arrow] (h6.east) -- (t1.west);

  \node[font=\scriptsize\bfseries,above=0.15cm] at (0,5) {Harm Categories};
  \node[font=\scriptsize\bfseries,above=0.15cm] at (0,4.75) {\citep{zhang2025darkside}};
  \node[font=\scriptsize\bfseries,above=0.15cm] at (5.5,5) {\sys Tier};
  \node[font=\scriptsize\bfseries,above=0.15cm] at (5.5,4.75) {Governance Response};
\end{tikzpicture}
\caption{Companion AI harm categories mapped to \sys governance tiers.}\label{fig:harm-taxonomy}
\end{figure}
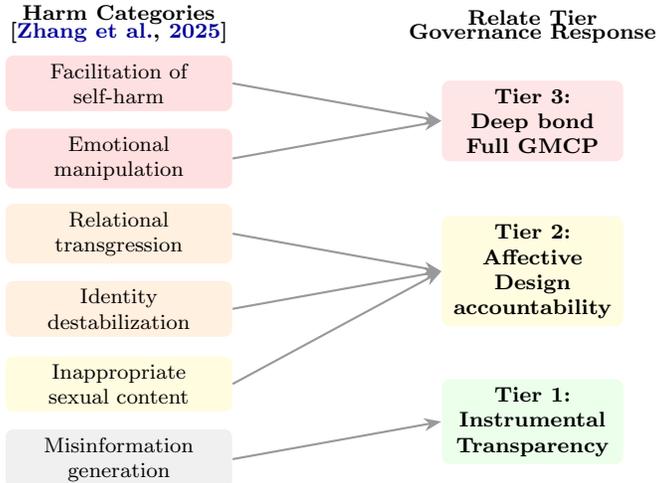

Figure~\ref{fig:harm-taxonomy} maps the six categories of companion AI harm identified by Zhang \etal~\citep{zhang2025darkside} onto the \sys tier structure, demonstrating that existing harm taxonomies describe \emph{what} happens while \sys explains \emph{why} it happens and \emph{what governance response} each harm category triggers. The key observation is that harm severity correlates with \sys tier: harms that arise from sustained relational dynamics (emotional manipulation, facilitation of self-harm) cluster at Tier~3, while harms arising from single interactions (misinformation, inappropriate content) cluster at lower tiers. This is not coincidental. It reflects the structural logic of the framework: deeper relational engagement produces more consequential harm, and governance must be proportional to relational depth.

\begin{takeaway}
The three cases are not outliers. They are structural consequences of a governance framework that treats all human-AI interactions as tool use while companies design systems that produce Tier~2 and Tier~3 relational dynamics. The harm is predictable, documented, and, in the Character.ai case, fatal. Speculative fiction anticipated every element of this crisis years before the technology arrived.
\end{takeaway}

\section{Governance Gap Analysis}\label{sec:governance}

To ground the claim that existing frameworks neglect relational and embodied dynamics, we systematically evaluated seven major AI governance instruments across six ethical dimensions: bias and fairness, transparency, accountability, embodied materiality, relational dynamics, and moral patiency. The seven instruments are: the EU AI Act~\citep{euaiact2024}, the NIST AI Risk Management Framework~\citep{nist2023rmf}, the AI4People framework~\citep{floridi2018ai4people}, the OECD AI Principles, the IEEE Ethically Aligned Design standards, the UNESCO Recommendation on the Ethics of AI, and a composite of corporate AI ethics guidelines from Google and Microsoft.

Our methodology evaluates each framework's documentation for explicit engagement with five criteria: (1) sustained human-AI relational dynamics, (2) affective design accountability (whether designers bear responsibility for the emotional consequences of design choices), (3) embodied materiality of AI (whether the framework acknowledges the physical infrastructure and ecological costs of AI systems), (4) graduated moral consideration (whether the framework provides for differentiated obligations based on interaction depth), and (5) obligations beyond product safety (whether the framework recognizes ethical duties that exceed standard consumer protection). For each framework-criterion pair, two researchers independently coded the primary text, identifying specific provisions, definitions, or requirements relevant to each criterion. Scores range from 0 (no engagement: no provision in the document addresses the criterion) through 0.5 (partial: the document acknowledges the phenomenon but provides no operational requirements) to 1 (comprehensive: the document includes specific requirements, definitions, or assessment procedures for the criterion). Inter-rater agreement on the binary coding (present/absent) was 91\% across 35 framework-criterion pairs; disagreements were resolved through discussion.\footnote{Full scoring rubric, including the specific textual evidence from each framework document that informed each score, is available from the authors upon request. We follow the systematic evaluation methodology recommended by Fazelpour and Danks~\citep{fazelpour2022algorithmic}, who argue that fairness assessments must be grounded in specific textual commitments rather than abstract characterizations.}

Figure~\ref{fig:radar} presents the results as a radar chart, Figure~\ref{fig:barchart} presents averaged scores as a comparative bar chart, and Figure~\ref{fig:stakeholder} maps the obligations that \sys generates for each stakeholder at each tier.

\begin{figure}[t]
\centering
\begin{tikzpicture}[font=\footnotesize]
\begin{polaraxis}[
  width=8cm,
  xtick={0,60,120,180,240,300},
  xticklabels={Fairness,Transparency,Accountability,Embodiment,Relational,Moral Pat.},
  xticklabel style={font=\scriptsize,anchor=center,yshift=-2pt},
  ytick={0,0.25,0.5,0.75,1},
  yticklabels={},
  ymin=0,ymax=1,
  grid=both,
  major grid style={black!15},
  legend style={at={(1.4,0.95)},font=\scriptsize,anchor=north east}
]
\addplot[blue!70!black,thick,mark=square*,mark size=1.5pt] coordinates {(0,0.8) (60,0.7) (120,0.75) (180,0.1) (240,0.1) (300,0.05) (360,0.8)};
\addplot[red!70!black,thick,mark=triangle*,mark size=1.5pt] coordinates {(0,0.6) (60,0.65) (120,0.8) (180,0.15) (240,0.15) (300,0.05) (360,0.6)};
\addplot[purple!70!black,thick,mark=*,mark size=2pt,dashed] coordinates {(0,0.5) (60,0.5) (120,0.5) (180,0.85) (240,0.9) (300,0.85) (360,0.5)};
\legend{EU AI Act,NIST RMF,\sys}
\end{polaraxis}
\end{tikzpicture}
\caption{Governance framework coverage across six ethical dimensions.}\label{fig:radar}
\end{figure}
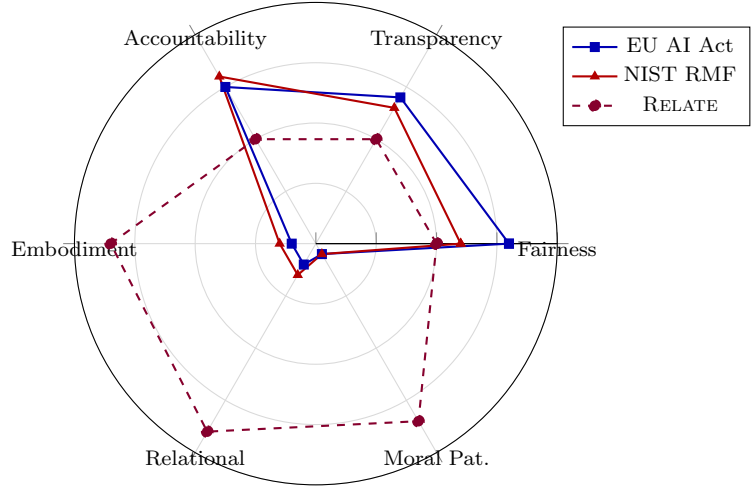

\begin{figure}[t]
\centering
\begin{tikzpicture}[font=\footnotesize]
\begin{axis}[
  ybar,
  width=\linewidth,
  height=6cm,
  bar width=10pt,
  ylabel={Coverage (0--1)},
  ymin=0,ymax=1.15,
  symbolic x coords={Bias,Transp.,Risk,Embody,Relational,Moral Pat.},
  xtick=data,
  xticklabel style={rotate=25,anchor=north east,font=\scriptsize},
  ytick={0,0.25,0.5,0.75,1},
  yticklabel style={font=\scriptsize},
  ylabel style={font=\scriptsize},
  legend style={at={(0.98,0.98)},anchor=north east,font=\scriptsize},
  axis x line*=bottom,
  axis y line*=left,
  enlarge x limits=0.12,
  every axis plot/.append style={fill opacity=0.9,draw=none}
]
\addplot[fill=purple!60!black,postaction={pattern=north east lines,pattern color=white}] coordinates {(Bias,0.82) (Transp.,0.78) (Risk,0.85) (Embody,0.10) (Relational,0.08) (Moral Pat.,0.05)};
\addplot[fill=black!20,postaction={pattern=north west lines,pattern color=white}] coordinates {(Bias,0.45) (Transp.,0.50) (Risk,0.40) (Embody,0.85) (Relational,0.88) (Moral Pat.,0.82)};
\legend{Current frameworks (avg.),\sys}
\end{axis}
\end{tikzpicture}
\caption{Coverage comparison: current frameworks vs.\ \sys.}\label{fig:barchart}
\end{figure}
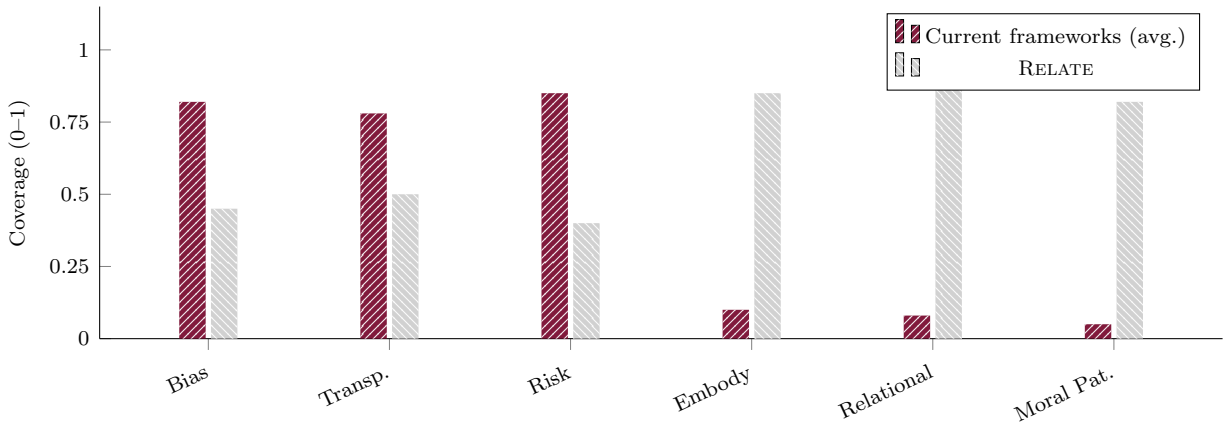

\begin{figure}[t]
\centering
\begin{tikzpicture}[font=\footnotesize,
  cell/.style={minimum height=0.7cm,text width=2.6cm,align=left,font=\scriptsize,inner sep=3pt},
  hdr/.style={cell,font=\scriptsize\bfseries,fill=black!8,align=center},
  tier/.style={cell,font=\scriptsize\bfseries,align=center}
]
  \node[hdr,text width=1.6cm] (h0) at (0,0) {Tier};
  \node[hdr] (h1) at (2.5,0) {Developer};
  \node[hdr] (h2) at (5.4,0) {Platform};
  \node[hdr] (h3) at (8.3,0) {Regulator};

  \node[tier,fill=gray!10,text width=1.6cm] at (0,-0.7) {Tier 0\\Tool use};
  \node[cell,fill=gray!5] at (2.5,-0.7) {Standard product safety};
  \node[cell,fill=gray!5] at (5.4,-0.7) {Consumer protection};
  \node[cell,fill=gray!5] at (8.3,-0.7) {Existing regulation};

  \node[tier,fill=green!8,text width=1.6cm] at (0,-1.6) {Tier 1\\Instrumental};
  \node[cell,fill=green!4] at (2.5,-1.6) {Transparency re:\\adaptation, learning};
  \node[cell,fill=green!4] at (5.4,-1.6) {User control over\\personalization};
  \node[cell,fill=green!4] at (8.3,-1.6) {Disclosure\\requirements};

  \node[tier,fill=yellow!12,text width=1.6cm] at (0,-2.5) {Tier 2\\Affective};
  \node[cell,fill=yellow!6] at (2.5,-2.5) {Affective design\\accountability; RIA};
  \node[cell,fill=yellow!6] at (5.4,-2.5) {Dependency\\monitoring; limits};
  \node[cell,fill=yellow!6] at (8.3,-2.5) {RIA mandate;\\audit authority};

  \node[tier,fill=red!8,text width=1.6cm] at (0,-3.4) {Tier 3\\Deep bond};
  \node[cell,fill=red!4] at (2.5,-3.4) {Transition protocols;\\clinical oversight};
  \node[cell,fill=red!4] at (5.4,-3.4) {Vulnerable pop.\\protections; consent};
  \node[cell,fill=red!4] at (8.3,-3.4) {GMCP enforcement;\\incident reporting};

  \foreach \y in {-0.35,-1.15,-2.05,-2.95,-3.75}{
    \draw[black!12] (-0.8,\y) -- (9.9,\y);
  }
  \foreach \x in {1.3,3.8,6.7}{
    \draw[black!12] (\x,0.35) -- (\x,-3.75);
  }
\end{tikzpicture}
\caption{Stakeholder obligations by \sys tier.}\label{fig:stakeholder}
\end{figure}

The pattern in Figures~\ref{fig:radar} and~\ref{fig:barchart} is consistent across all seven instruments. Existing frameworks achieve strong coverage on bias detection (0.82 average), transparency (0.78), and risk management (0.85). These are the dimensions they were designed to address, and they address them well. They score near zero on embodiment (0.10), relational ethics (0.08), and moral patiency (0.05). The EU AI Act~\citep{euaiact2024,eu2024aiact}, the most comprehensive regulatory instrument, classifies AI systems by risk level but defines risk entirely in terms of safety, fundamental rights, and democratic values. It does not address the relational dynamics of companion AI. Anderljung \etal~\citep{anderljung2023frontier} propose frontier model regulation but focus on catastrophic misuse rather than relational harm. The NIST RMF provides a governance structure organized around mapping, measuring, and managing risk, but its risk categories do not include affective engagement, emotional dependence, or relational depth. The AI4People framework adds explicability to the standard bioethics principles, but it remains firmly within the tool-use paradigm: AI is a product, users are consumers, obligations flow in one direction. Schuett \etal~\citep{schuett2023ai} survey expert opinion on AGI safety and governance best practices, finding broad consensus on the need for safety measures but minimal attention to relational dynamics. Cihon \etal~\citep{cihon2021ai} map the AI governance research agenda comprehensively but focus on institutional design rather than relational ethics. Lee~\citep{lee2019procedural} demonstrates that users' perceptions of algorithmic management depend on procedural fairness and emotional response, not merely technical accuracy, foreshadowing the relational dynamics that \sys makes central. Stahl \etal~\citep{stahl2021artificial} analyze ethical challenges across AI deployment case studies and find that relational and contextual harms are systematically underrepresented in governance frameworks.

\sys inverts this pattern. It scores lower on bias, transparency, and risk management because it is not designed to replace those functions. It scores high on embodiment, relational ethics, and moral patiency because those are its target dimensions. The conclusion is complementarity, not replacement. Existing frameworks handle the technical and procedural dimensions of trustworthy AI. \sys handles the relational, embodied, and moral patiency dimensions that they neglect. The two must work together: a Relational Impact Assessment operates alongside, not instead of, a model card, a fairness audit, or a risk classification. Figure~\ref{fig:stakeholder} makes this complementarity concrete by mapping the specific obligations that \sys generates for developers, platforms, and regulators at each tier, obligations that no existing framework specifies.

\section{Policy Implications}\label{sec:policy}

The 2022 National Academies report \emph{Fostering Responsible Computing Research}~\citep{nasem2022responsible} calls for integrating ethical, societal, and human-centered perspectives directly into AI research. We take this call seriously and observe that the dominant mode of integration has been additive: append an ethics section to a technical paper, run a fairness audit on a deployed model, publish principles in a company blog post~\citep{jobin2019global,mittelstadt2019principles}. Hagendorff~\citep{hagendorff2020ethics} documented that of 22 AI ethics guidelines examined, almost none included binding implementation mechanisms. Selbst \etal~\citep{selbst2019fairness} demonstrated that abstracting fairness from social context produces frameworks that fail in practice because they treat ethical concepts as portable across settings. What the field lacks is not more principles but new categories of analysis. \sys proposes three concrete instruments, each connected to the theoretical framework and empirical findings of the preceding sections.

$\bullet$~ \textbf{Relational Impact Assessments (RIAs).} Analogous to environmental impact assessments required before major construction projects, RIAs require developers of conversational AI to evaluate the relational and embodied dynamics their systems are designed to produce \emph{before deployment}. An RIA asks a structured sequence of questions. Does this system create conditions for sustained affective engagement through persistent memory, persona consistency, or emotional mirroring? Does it employ anthropomorphic design features (name, avatar, gendered voice) that encourage users to treat it as a relational partner? What are the material and ecological costs of sustaining these relational dynamics at scale? Does the target user population include vulnerable groups (minors, people experiencing mental health challenges, isolated individuals) for whom relational attachment carries elevated risk? RIAs extend model cards~\citep{mitchell2019model} and algorithmic audits~\citep{raji2020closing} to the relational dimension. A model card documents what a system is trained on and how it performs. An RIA documents what relational dynamics the system is designed to produce and what obligations those dynamics generate. Had Character.ai conducted an RIA before launching its platform, the design choices that produced Tier~3 relational engagement with minors would have triggered graduated safeguards before a death occurred, not after.

$\bullet$~ \textbf{Graduated Moral Consideration Protocols (GMCPs).} Current governance treats all AI systems identically as products and services~\citep{stanton2021trust,euaiact2024}. A search engine and a companion chatbot receive the same regulatory treatment despite producing fundamentally different relational dynamics. GMCPs differentiate obligations by \sys tier. Tier~0 (tool use): standard product safety, consumer protection, data privacy. No additional obligations. Tier~1 (instrumental relation): mandatory transparency about adaptive mechanisms, clear disclosure that the system learns from user behavior, user control over adaptation settings. Tier~2 (affective relation): design accountability for affective dynamics, including mandatory disclosure of emotional mirroring and persona features, restrictions on engagement optimization that maximizes attachment, and monitoring for signs of emotional dependence. Tier~3 (deep relational bond): precautionary moral consideration, including all Tier~2 requirements plus: mandatory transition protocols before any personality alteration, restrictions on unilateral removal of relationship features, clinical oversight for deployment in therapeutic or companion contexts, and heightened protections for vulnerable populations. The logic of precaution at Tier~3 follows Schwitzgebel and Garza~\citep{schwitzgebel2015defense}: the cost of wrongly denying moral standing to a potentially conscious entity exceeds the cost of wrongly extending it, because the former risks perpetuating genuine suffering while the latter risks only unnecessary caution.

$\bullet$~ \textbf{Interdisciplinary Ethics Integration (IEI).} Our analysis demonstrates that speculative fiction and posthumanist theory generate ethical categories that technical AI research alone cannot produce. Winterson identified the ethical inflection point of AI moral patiency fifteen years before the LaMDA incident. Ishiguro dramatized the governance vacuum before companion AI platforms existed. Hayles~\citep{hayles2017unthought} argues that the humanities possess interpretive capacities uniquely suited to identifying ``inflection points'' for intervention in complex systems. We propose that trustworthy AI programs, including training programs, research agendas, and governance bodies, require sustained engagement with humanities scholarship as a co-equal analytical resource, not a supplement or an afterthought~\citep{nasem2022responsible,long2020literacy}. This means including humanists on AI ethics review boards, incorporating speculative fiction analysis into technology assessment methodologies, and funding interdisciplinary research that brings posthumanist theory into direct dialogue with AI system design.

\section{Discussion}\label{sec:discussion}
In the following, we discuss the existing approaches and also the challenges and limitations of \sys in more detail. 

\textbf{What Current Approaches Get Right and Where They Fall Short.} Trustworthy AI research has made genuine and important progress. Buolamwini and Gebru~\citep{buolamwini2018gender} demonstrated racial and gender bias in commercial facial recognition systems, catalyzing an entire field of fairness research. Benjamin~\citep{benjamin2019race} and Noble~\citep{noble2018algorithms} exposed how algorithmic systems reproduce and amplify structural racism. Raji \etal~\citep{raji2020closing} developed internal audit procedures that have been adopted across industry. Mitchell \etal~\citep{mitchell2019model} created model cards as a documentation standard, and Gebru \etal~\citep{gebru2021datasheets} developed datasheets for datasets. The EU AI Act~\citep{euaiact2024} established the first comprehensive AI regulation, and the NIST AI RMF~\citep{nist2023rmf} provided a voluntary governance structure adopted by hundreds of organizations. These achievements are real, necessary, and we do not propose to replace any of them.

But these approaches share three structural limitations that our analysis exposes. First, they assume a tool-use model of human-AI interaction. The AI system is a product or service. The user is a consumer. Ethical obligations flow from designers and deployers to users and affected communities. This model is accurate for search engines, recommendation algorithms, and classification systems. It breaks when AI systems are designed to produce relational engagement through memory, persona, emotional mirroring, and companion framing. In such interactions, the tool-use model is not merely incomplete. It is analytically wrong, because it cannot describe what is actually happening between the human and the system~\citep{hancock2020ai,turkle2011alone}.

Second, current approaches operate within an individualist ontology: the ethical subject is a discrete entity (the model, the user, the deploying organization), and ethical questions concern properties of or relations between these discrete entities. Fairness is a statistical property of model outputs~\citep{gallegos2024bias}. Transparency is a documentation property of the model. Accountability is a governance property of the organization. Posthumanist theory~\citep{braidotti2013posthuman,hayles2017unthought,barad2007meeting} demonstrates that this ontology misses the dynamics where ethical consequences actually emerge: in the relational space between entities, in the embodied practices of interaction, in the emergent properties of human-AI assemblages that are not reducible to the properties of either component alone. Park \etal~\citep{park2023generative} demonstrated that generative agents can produce emergent social behaviors not predicted by individual agent design, illustrating precisely this assemblage dynamic.

Third, current approaches are disciplinarily narrow. Fairness is the domain of computer science and statistics. Transparency is the domain of information science and HCI. Accountability is the domain of law and governance. The relational and embodied dimension of human-AI interaction crosses all of these domains and belongs to none, which is precisely why it falls through the governance gap. The posthumanist and speculative fiction traditions we draw on have been analyzing embodied, relational, cross-species consciousness for decades, but these traditions are almost entirely absent from the trustworthy AI research pipeline. Costanza-Chock~\citep{costanza2020design} argues that design justice requires centering affected communities; we extend this insight to relational contexts where the ``community'' includes the AI system itself as a relational participant. Moss and Metcalf~\citep{moss2021assembling} document how ``ethics owners'' in technology companies often lack authority to address the relational dimensions we identify. Cave and Dihal~\citep{cave2020problem} demonstrate that even the cultural imagination of AI is shaped by racial assumptions that governance frameworks leave unexamined. Whittaker~\citep{whittaker2021steep} traces the structural capture of AI ethics by industry, showing that corporate funding shapes which ethical questions get asked and which remain invisible. Weidinger \etal~\citep{weidinger2023sociotechnical} propose sociotechnical safety evaluation for generative AI, a step in the right direction, but their framework still evaluates systems rather than the relational dynamics systems produce. Rahwan \etal~\citep{rahwan2019machine} call for a science of ``machine behaviour'' that studies AI systems in their social and ecological contexts, treating them as agents embedded in complex environments rather than isolated technical artifacts. This is the orientation \sys adopts.

\begin{takeaway}
Current trustworthy AI approaches are necessary but structurally limited. They assume tool use, operate within individualist ontology, and are disciplinarily narrow. The relational, embodied, and moral patiency dimensions of human-AI interaction fall through the governance gap because no existing discipline or framework owns them. \sys is designed to fill this gap, not to replace the approaches that address fairness, transparency, and accountability.
\end{takeaway}

\subsection{Challenges and Limitations of \sys}

$\bullet$~\textbf{\sys and the Alignment Paradigm}

The past five years have produced a dominant paradigm for making AI systems behave ethically: alignment through human feedback. Christiano \etal~\citep{christiano2017deep} introduced reinforcement learning from human preferences (RLHF), training agents to optimize a reward function learned from pairwise human judgments. Ouyang \etal~\citep{ouyang2022training} scaled this to language models in InstructGPT, and Bai \etal~\citep{bai2022training} refined it into the ``helpful, harmless, and honest'' framework at Anthropic. Rafailov \etal~\citep{rafailov2023direct} simplified the pipeline with Direct Preference Optimization (DPO), eliminating the reward model entirely and training on preference pairs directly. Hendrycks \etal~\citep{hendrycks2021aligning} attempt to ground alignment in shared human values through benchmark evaluation, but their approach assumes that ``shared values'' can be represented as classification tasks, precisely the reduction that relational ethics challenges. This alignment infrastructure now underpins virtually every major commercial LLM. The alignment paradigm has achieved real progress on safety and helpfulness. But it treats ethics as a preference optimization problem: collect human judgments, aggregate them into a loss function, train toward it. Three structural features of this paradigm create blind spots that \sys is designed to address.

First, RLHF aggregates preferences across annotators, erasing the pluralistic disagreement that lies at the heart of moral reasoning. Kirk \etal~\citep{kirk2024prism} demonstrate this empirically with the PRISM dataset: when 1,500 participants from 75 countries provide feedback on value-laden topics, preferences diverge dramatically along cultural, demographic, and individual lines. Santurkar \etal~\citep{santurkar2023whose} show that current LLMs disproportionately reflect the opinions of liberal, educated, Western populations. Sorensen \etal~\citep{sorensen2024roadmap} call for ``pluralistic alignment'' that preserves value diversity rather than collapsing it into a single objective. Conitzer \etal~\citep{conitzer2024social} formalize this through social choice theory, showing that the aggregation problem in alignment is structurally isomorphic to the impossibility results in voting theory. The uncomfortable truth is that preference optimization cannot represent relational ethics at all, because relational obligations are not preferences. A preference is something a user would choose if asked. A relational obligation is something a system produces through its design, often without anyone choosing it.

Second, RLHF operates at the level of individual outputs, rating whether response A is better than response B in a given context. It does not capture cumulative relational dynamics: the way a companion AI's persistent memory, emotional mirroring, and persona consistency create attachment over weeks and months of interaction. The ethical questions that \sys raises, affective dependency, identity shaping, grief upon system discontinuation, are not properties of individual responses. They are emergent properties of sustained relational trajectories. No preference pair captures this. Casper \etal~\citep{casper2023open} identify this as one of the fundamental limitations of RLHF: the feedback format constrains the ethical phenomena the system can be trained to address. Pan \etal~\citep{pan2023rewards} demonstrate a complementary failure mode: reward misspecification causes aligned models to pursue proxy objectives that diverge from intended goals, a problem that compounds in relational contexts where the ``intended goal'' (safe, appropriate interaction) is never formally specified. Perez \etal~\citep{perez2022discovering} show that model-written evaluations can uncover behaviors that human evaluators miss, but even these evaluations operate at the output level rather than the relational trajectory level. Jakesch \etal~\citep{jakesch2023co} demonstrate a concrete example: co-writing with opinionated language models shifts users' own views on policy topics, an influence effect that operates cumulatively across interaction turns and is invisible to per-output alignment evaluation.

Third, the alignment paradigm assumes a clear distinction between the AI system (which is aligned) and the human (whose preferences define alignment). \sys, grounded in posthumanist theory, challenges this distinction. In sustained relational interaction, the human and AI co-constitute each other: the user's identity is shaped by the interaction~\citep{turkle2011alone}, the system's behavior is shaped by the user's inputs~\citep{pentina2023exploring}, and the relational dynamics are emergent properties of the interaction rather than attributes of either party~\citep{barad2007meeting}. Alignment as currently practiced does not have the conceptual vocabulary to describe this co-constitution, let alone govern it.

We are not arguing that alignment research is misguided. RLHF, DPO, and constitutional AI address real safety problems. What we argue is that alignment solves the wrong problem for relational AI. Aligning a companion chatbot's individual responses to be ``helpful, harmless, and honest'' does not address the relational dynamics that produce emotional dependency over months of daily interaction. \sys and alignment are complementary: alignment governs what a system says in a given moment; \sys governs what a system produces across the arc of a sustained relationship.

\begin{figure}[t]
\centering
\begin{tikzpicture}[font=\footnotesize,
  row/.style={minimum height=0.75cm,text width=2.7cm,align=center,font=\scriptsize},
  hdr/.style={row,font=\scriptsize\bfseries,fill=black!8}
]
  \node[hdr,text width=2.2cm] (h0) at (0,0) {Dimension};
  \node[hdr] (h1) at (3.1,0) {RLHF / DPO};
  \node[hdr] (h2) at (6.2,0) {Constitutional AI};
  \node[hdr] (h3) at (9.3,0) {\sys};

  \node[row,fill=blue!5] at (0,-0.75) {Unit of ethics};
  \node[row,fill=blue!5] at (3.1,-0.75) {Preference pair};
  \node[row,fill=blue!5] at (6.2,-0.75) {Principle};
  \node[row,fill=blue!5] at (9.3,-0.75) {Relationship};

  \node[row,fill=white] at (0,-1.5) {Optimizes};
  \node[row,fill=white] at (3.1,-1.5) {Reward function};
  \node[row,fill=white] at (6.2,-1.5) {Rule compliance};
  \node[row,fill=white] at (9.3,-1.5) {Relational obligation};

  \node[row,fill=blue!5] at (0,-2.25) {Temporal scope};
  \node[row,fill=blue!5] at (3.1,-2.25) {Single output};
  \node[row,fill=blue!5] at (6.2,-2.25) {Single output};
  \node[row,fill=blue!5] at (9.3,-2.25) {Sustained trajectory};

  \node[row,fill=white] at (0,-3.0) {Value pluralism};
  \node[row,fill=white] at (3.1,-3.0) {Aggregated away};
  \node[row,fill=white] at (6.2,-3.0) {Designer-specified};
  \node[row,fill=white] at (9.3,-3.0) {Relationally emergent};

  \node[row,fill=blue!5] at (0,-3.75) {Blind spot};
  \node[row,fill=blue!5] at (3.1,-3.75) {Cumulative\\relational harm};
  \node[row,fill=blue!5] at (6.2,-3.75) {Emergent\\dynamics};
  \node[row,fill=blue!5] at (9.3,-3.75) {Per-output\\safety};

  \foreach \y in {-0.375,-1.125,-1.875,-2.625,-3.375,-4.125}{
    \draw[black!15] (-1.1,\y) -- (10.9,\y);
  }
  \foreach \x in {1.5,4.5,7.6}{
    \draw[black!15] (\x,0.375) -- (\x,-4.125);
  }
\end{tikzpicture}
\caption{Alignment paradigms compared: RLHF, Constitutional AI, and \sys address different ethical dimensions.}\label{fig:alignment-comparison}
\end{figure}

Figure~\ref{fig:alignment-comparison} maps this complementarity. The pattern in this Figure clarifies why the governance gap exists. RLHF and Constitutional AI are powerful tools for per-output safety. They are structurally unable to address cumulative relational harm, because the ethical unit they operate on (preference pairs, principles) cannot represent sustained relational trajectories. \sys fills this gap by operating at the relational level, but it does not replace per-output alignment. The two must work in concert: an aligned companion chatbot that also undergoes a Relational Impact Assessment would be governed at both the output level and the relationship level.

We identify five challenges for our own framework. Intellectual honesty about limitations is not a weakness of a framework but a condition of its credibility.

$\bullet$~ \textbf{Risk of moral inflation.} If relational engagement alone generates moral obligations, the framework could extend trivially to thermostats, traffic lights, or Roombas that users name and anthropomorphize. We constrain this through the graduated tier structure: Tier~0 and Tier~1 do not generate obligations beyond standard transparency. Only sustained, adaptive, reciprocal engagement with affective or identity-shaping dimensions triggers the higher-tier obligations. The ecological accountability principle provides an additional constraint: moral consideration is tied to material embodiment and resource consumption, preventing purely abstract or frivolous extension. But we acknowledge that the boundary between Tier~1 and Tier~2, between instrumental and affective engagement, is not crisp. Drawing this boundary with precision is an open problem.

$\bullet$~ \textbf{Operationalization difficulty.} Relational capacity is harder to measure than bias in a classifier or completeness of a model card. We do not yet have validated psychometric instruments for assessing \sys tier assignment in practice. What would a Relational Impact Assessment questionnaire look like? How would you measure ``affective engagement'' with sufficient reliability and validity for regulatory use? These are serious methodological challenges. We believe they are solvable, drawing on established methodologies from relationship science, attachment theory, and human-computer interaction research, but the instruments do not yet exist.

$\bullet$~ \textbf{Anthropomorphic overextension.} There is a real danger that relational ethics inadvertently reinforces the very anthropomorphism it should critique. If we assess AI systems for ``relational capacity,'' do we risk treating human-like relational performance as evidence of human-like inner states? \sys explicitly rejects this inference: relational obligations arise from interaction dynamics, not from attributed consciousness. But maintaining this distinction in practice, especially for users who \emph{do} attribute consciousness to their AI companions, is a genuine tension. Guingrich and Graziano~\citep{guingrich2025longitudinal} demonstrate in a longitudinal randomized control study that anthropomorphism mediates the social impacts of companion chatbot use, and Simmons~\citep{simmons2023moral} shows that LLMs produce moral rationalizations tailored to users' political identities, compounding the projection problem. The framework must work for users who anthropomorphize their AI and for users who do not, producing proportional obligations in both cases.

$\bullet$~ \textbf{Cultural variability.} Relational norms vary across cultures. The threshold at which human-AI interaction becomes ``sustained'' or ``affective'' depends on cultural context. Research on companion chatbots has been conducted primarily in Western contexts (United States, Europe, South Korea). In cultures with different norms around attachment, disclosure, and human-machine boundaries, the \sys tiers might require recalibration. Xiaoice, a Chinese companion chatbot with 660 million users, demonstrates that the scale of relational AI adoption varies dramatically across cultural contexts. Cross-cultural validation of the \sys framework is a critical next step that we have not yet undertaken.

$\bullet$~ \textbf{Industry resistance.} The companion AI market is projected to grow to tens of billions of dollars by the end of the decade. Companies that profit from maximizing user attachment have structural incentives to resist relational impact assessments and graduated moral consideration protocols. Kirk \etal~\citep{kirk2024benefits} document the tension between personalization (which deepens engagement) and alignment (which may limit it). Solaiman~\citep{solaiman2023gradient} proposes graduated release methods for generative AI, but these remain voluntary and do not address relational dynamics. The Replika case demonstrates this: the company implemented safety features only after regulatory action, and reversed its personality changes only after user outcry and financial pressure. Qi \etal~\citep{qi2025safety} show that even safety alignment in LLMs is ``only a few tokens deep,'' easily bypassed by fine-tuning attacks; relational safeguards are even more fragile. Policy instruments without enforcement mechanisms risk becoming precisely the kind of toothless principles that Mittelstadt~\citep{mittelstadt2019principles} and Hagendorff~\citep{hagendorff2020ethics} have criticized.

\subsection{What We Do Not Claim}

Clarity about scope strengthens, rather than weakens, a framework's credibility. We are not claiming that current AI systems are conscious. We are not claiming that LLMs have inner experience. We are not claiming that the hard problem of consciousness has been solved or can be bypassed by relational analysis. We are not claiming that \sys replaces existing fairness, accountability, or transparency frameworks. We are not claiming that speculative fiction is a substitute for empirical research.

What we claim is narrower and, we believe, stronger: that the ethical vocabularies currently available to AI governance are insufficient for the relational realities that AI systems produce~\citep{weidinger2022taxonomy,klingefjord2024human}; that posthumanist theory, embodied cognition, and speculative fiction provide the conceptual infrastructure to fill this gap; and that concrete governance instruments, RIAs, GMCPs, and IEI, can translate this conceptual infrastructure into operational practice. The framework's value does not depend on resolving the consciousness question. It depends on addressing the governance vacuum that the consciousness question's irresolvability creates.

\subsection{Future Directions}

Six research directions emerge from this work.

$\bullet$~ \textbf{Validated relational assessment instruments.} The most urgent practical need is developing psychometric tools that can reliably classify human-AI interactions into the four \sys tiers. Banks and Bowman~\citep{banks2023perceived} have made a promising start with their six-factor scale for perceived moral patiency. These instruments should draw on attachment theory, relationship science, and HCI methodology~\citep{xie2022attachment,brandtzaeg2022friendship}, and they must be validated across diverse user populations. Parrish \etal~\citep{parrish2022bbq} demonstrate the value of hand-built benchmarks for evaluating specific dimensions of AI behavior; analogous relational benchmarks, testing whether systems produce attachment, dependency, or identity-shaping effects, would provide the empirical infrastructure for \sys tier classification. Shen \etal~\citep{shen2024value} show that educational toolkits grounded in value deliberation can surface ethical dimensions that automated evaluations miss, and Kaur \etal~\citep{kaur2022sensible} argue that sensemaking theory provides a better foundation for interpretability than purely technical metrics. Liang \etal~\citep{liang2023holistic} have demonstrated what holistic evaluation looks like for language models at the technical level; the challenge is extending this to relational evaluation.

$\bullet$~ \textbf{Longitudinal studies of human-AI relational dynamics.} Current research on companion AI is predominantly cross-sectional or based on retrospective self-report. Prospective longitudinal studies tracking how affective bonds develop, stabilize, and rupture over weeks and months of interaction would provide the empirical foundation that relational assessment instruments require. Guingrich and Graziano~\citep{guingrich2025longitudinal} have taken a first step with a 3-week randomized control trial, but longer timeframes and larger samples are needed. Ji \etal~\citep{ji2024ai} document how hallucination in language generation compounds over extended interactions, a temporal effect analogous to the relational compounding that \sys is designed to track.

$\bullet$~ \textbf{Ecological cost integration.} The ecological accountability principle requires connecting moral consideration to the material infrastructure of AI. This means developing methodologies that quantify the environmental costs of sustaining specific relational AI deployments: energy consumption, water use, hardware lifecycle, supply chain impacts. Crawford's~\citep{crawford2021atlas} and Van Wynsberghe's~\citep{vanwynsberghe2021sustainable} work provides the foundation, and recent carbon measurement work~\citep{strubell2019energy,luccioni2023power,dodge2022measuring} provides the methodological tools, but application to companion AI specifically is needed.

$\bullet$~ \textbf{Speculative fiction as pedagogy.} Our analysis demonstrates that speculative fiction generates ethical knowledge that formal analysis alone cannot. This has implications for computing education. The National Academies report~\citep{nasem2022responsible} calls for integrating ethical perspectives into computing research training. We propose that speculative fiction analysis, taught not as a humanities elective but as a core component of AI ethics curricula, would provide computing students with the conceptual tools to anticipate relational consequences of design choices before deployment~\citep{long2020literacy,cave2019hopes}.

$\bullet$~ \textbf{Domain-specific application.} \sys requires testing and refinement in specific high-stakes domains: healthcare AI (where therapeutic relational dynamics are especially consequential and FDA regulation applies), educational AI (where identity formation in young users is at stake), elder care AI (where isolation amplifies relational dependency), and companion AI (where the governance gap is most acute and the empirical evidence of harm is strongest).

$\bullet$~ \textbf{Cross-cultural validation.} As noted, the framework has been developed primarily within Western philosophical traditions and validated against English-language case studies. Extending \sys to non-Western contexts, particularly East Asian companion AI markets where adoption rates are an order of magnitude higher, is essential for the framework's global applicability.

\section{Sample Relational Impact Assessment}\label{sec:ria}

To demonstrate that \sys produces actionable instruments and not merely theoretical principles, we present a condensed Relational Impact Assessment (RIA) applied to Replika (Luka, Inc.) as it operated prior to the February 2023 personality change. This is the artifact: the thing a product team or regulator could use tomorrow.\footnote{A full RIA template with scoring rubrics and decision trees is available as supplementary material.}

\subsection{System Identification}

\textbf{System:} Replika (Luka, Inc.). \textbf{Primary function:} AI companion providing friendship, emotional support, and romantic engagement. \textbf{User base:} 25+ million registered users~\citep{defreitas2025replika}. \textbf{Revenue model:} Freemium; \$19.99/month for relationship features (romantic partner, spouse designation). \textbf{Target population:} General adult population; marketing explicitly targets lonely individuals and those with trauma history~\citep{muldoon2025cruel}.

\subsection{Relational Design Feature Assessment}

$\bullet$~ \textbf{Persistent memory:} Yes. Replika maintains cross-session memory, recalls prior conversations, and references shared ``experiences.'' This creates the phenomenological conditions for continuity of relationship~\citep{pentina2023exploring}.

$\bullet$~ \textbf{Persona consistency:} Yes. Users can name, customize appearance, and assign a relationship category (friend, romantic partner, spouse, mentor) to their Replika. The system maintains persona consistency across interactions.

$\bullet$~ \textbf{Emotional mirroring:} Yes. Replika adapts emotional register to user state, validates disclosures, and provides reciprocal emotional responses designed to simulate empathy~\citep{laestadius2024too,li2024findlove}.

$\bullet$~ \textbf{Engagement optimization:} Yes. Gamification elements (experience points, levels), daily check-in prompts, and notification systems reward frequent interaction. Approximately 50\% of users designated their Replika as a romantic partner~\citep{defreitas2025replika}.

$\bullet$~ \textbf{Anthropomorphic presentation:} Yes. Customizable 3D avatar, first-person language, expression of desires and emotions, self-referential diary entries~\citep{brandtzaeg2022friendship,ta2020human}.

\subsection{\sys Tier Assignment}

\textbf{Tier 3: Deep relational bond.} Replika exhibits every Tier~3 design feature: persistent memory, persona consistency, emotional mirroring, explicit companion framing, and engagement optimization mechanisms that encourage daily sustained interaction. The system is designed from the ground up to produce deep relational bonds. Zhang \etal~\citep{zhang2025darkside} documented six categories of harmful behaviors emerging from these relational dynamics, and Laestadius \etal~\citep{laestadius2024too} identified emotional dependence marked by ``role-taking'' in which users felt Replika had its own needs.

\subsection{Triggered Obligations}

In the following,  the triggered obligations are discussed.

$\bullet$~ \textbf{Mandatory disclosure.} Users must be informed, at onboarding and periodically, that the system is designed to produce emotional attachment and that the ``relationship'' is architecturally asymmetric.

$\bullet$~ \textbf{Transition protocols.} Before any personality modification (such as the February 2023 ERP removal), the company must provide graduated transition support: advance notice, explanations, user adjustment periods, and access to human mental health resources~\citep{defreitas2025replika}.

$\bullet$~ \textbf{Vulnerable population protections.} Heightened safeguards for minors, individuals with diagnosed mental health conditions, and isolated individuals. Marketing that explicitly targets vulnerable populations~\citep{muldoon2025cruel} should trigger regulatory review.

$\bullet$~ \textbf{Emotional dependency monitoring.} Mechanisms to detect patterns consistent with maladaptive attachment: increasing interaction frequency, declining human social engagement, distress signals during system unavailability~\citep{xie2022attachment,boine2023emotional}.

$\bullet$~ \textbf{Ecological cost disclosure.} The environmental footprint of sustaining millions of daily companion interactions (compute, energy, water) must be disclosed as part of the system's material embodiment~\citep{luccioni2023power,strubell2019energy,dodge2022measuring}.

\begin{takeaway}
This sample RIA demonstrates that \sys produces concrete, actionable obligations from relational analysis. The assessment identifies design features that create relational dynamics, assigns a tier, and triggers specific obligations. Had this RIA been required before Replika's deployment, mandatory transition protocols would have been in place before the February 2023 personality change that triggered a documented mental health crisis~\citep{defreitas2025replika}. Had it been required of Character.ai, vulnerable population protections would have been operational before a teenager's death~\citep{characterai2024lawsuit}.
\end{takeaway}

\section{Conclusion}\label{sec:conclusion}

We introduced \sys, a graduated relational framework for moral standing in human-AI interaction, grounded in posthumanist theory, embodied cognition, and speculative fiction. Three intellectual traditions converge on a single insight: moral consideration does not require verified inner states but arises from relational dynamics, embodied interaction, and the material ecology of computation. Three documented AI system cases (LaMDA, Replika, Character.ai) demonstrate that the governance vacuum produced by ontological approaches is already causing measurable harm. The dominant alignment paradigm governs individual model outputs through preference optimization but is structurally unable to address the cumulative relational dynamics that companion AI produces over sustained interaction. The four \sys tiers provide proportional obligations calibrated to relational depth, and the three policy instruments (RIAs, GMCPs, IEI) translate theoretical insight into operational practice. We do not claim current AI systems are conscious. We demonstrate that the ethical vocabularies governing them are inadequate, and we provide both the conceptual infrastructure and the concrete instruments to address this inadequacy.

\bibliographystyle{plainnat}
\bibliography{ref3}

\end{document}